\documentclass[prd,11pt]{article}

\usepackage[utf8]{inputenc} % allow utf-8 input
\usepackage[T1]{fontenc}    % use 8-bit T1 fonts
\usepackage{hyperref}       % hyperlinks
\usepackage{url}            % simple URL typesetting
\usepackage{booktabs}       % professional-quality tables
\usepackage{amsfonts}       % blackboard math symbols
\usepackage{nicefrac}       % compact symbols for 1/2, etc.
\usepackage{microtype}      % microtypography
\usepackage{geometry}
\usepackage{authblk}
\usepackage{latexsym}

\usepackage{enumerate}
\usepackage[inline]{enumitem}
\usepackage{amsmath,amssymb}
\usepackage{amsfonts,dsfont}
\usepackage{nicefrac}
\usepackage{microtype}
\usepackage{mathtools}

\usepackage{sidecap}
\usepackage{caption}
\usepackage{subcaption}
\usepackage{wrapfig}
\usepackage{enumitem}
\usepackage{algorithm}
\usepackage[noend]{algorithmic}
\usepackage[normalem]{ulem}
\usepackage{amssymb}
\usepackage{multicol}
\usepackage{adjustbox}
\usepackage{multirow}
\usepackage{color}
\usepackage{xspace}
\usepackage{CJKutf8}

\PassOptionsToPackage{numbers}{natbib}
\usepackage{natbib}

\usepackage{graphicx}
\usepackage{booktabs} % for professional tables
\usepackage{mathtools}

\usepackage{hyperref}

\usepackage{math_definitions}
\newcommand{\dataset}{X}
\newcommand{\datasetcomplement}[1]{\dataset \backslash #1}

\newcommand{\tree}{\mathtt{H}}
\newcommand{\treeset}{\mathcal{H}}

\newcommand{\Efun}{\ensuremath{\phi}}

\newcommand{\EfunS}{\ensuremath{\psi}}
\newcommand{\EfunT}{\ensuremath{\phi}}

\newcommand{\treestar}{\ensuremath{\tree^\star}}
\newcommand{\NumHierachicalClusters}{\omega(N  N! / 2^{N-1})}
\newcommand{\exactNumClusterings}{(2N-3)!!}

\newcommand{\trellisVertex}{\mathbb{V}}
\newcommand{\children}{\mathbb{C}}
\renewcommand{\cite}{\citep}

\newcount\Comments  % 0 suppresses notes to selves in text
\definecolor{darkgreen}{rgb}{0,0.5,0}
\definecolor{darkred}{rgb}{0.7,0,0}
\definecolor{teal}{rgb}{0.1,0.6,0.7}
\definecolor{blue}{rgb}{0.0,0.1,0.9}
\definecolor{orange}{rgb}{1.,0.7,0.0}
\definecolor{lightblue}{rgb}{0.70, 0.80, 0.89}
\newcommand{\kibitz}[2]{\ifnum\Comments=1{{\textcolor{#1}{\textsf{\footnotesize [#2]}}}}\fi}

\newcommand\blfootnote[1]{%
  \begingroup
  \renewcommand\thefootnote{}\footnote{#1}%
  \addtocounter{footnote}{-1}%
  \endgroup
}

\def\beq{\begin{equation}}
\def\eeq{\end{equation}}
\newcommand{\bea}{\begin{eqnarray}\begin{aligned}}
\newcommand{\eea}{\end{aligned}\end{eqnarray}}
\def\bitem{\begin{itemize}}
\def\eitem{\end{itemize}}

\vspace{1cm}
\author[1,2]{Craig S. Greenberg\thanks{Both authors contributed equally to this work.}$^,$}
\author[3]{Sebastian Macaluso$^{\ast, }$}
\author[1]{Nicholas Monath}
\author[4]{Ji-Ah Lee}
\author[4]{Patrick  Flaherty}
\author[3]{Kyle Cranmer}
\author[1]{Andrew McGregor}
\author[1]{Andrew McCallum}

\date{} 
\affil[1]{\small{College of Information and Computer Sciences, University of Massachusetts Amherst, USA}}
\affil[2]{\small{National Institute of Standards and Technology, USA}}
\affil[3]{\small{Center for Cosmology and Particle Physics \& Center for Data Science, New York University, USA}}
\affil[4]{\small{Department of Mathematics and Statistics, University of Massachusetts Amherst, USA}}

\providecommand{\keywords}[1]{\textit{Keywords:} #1}

\begin{document}
\title{Data Structures \& Algorithms for Exact Inference \\in Hierarchical Clustering}

\vskip 0.3in

\maketitle
\vspace{-1cm}
\begin{center}
\end{center}
\vspace{0.5cm}
\begin{abstract}
\vspace{-0.2cm}
\noindent Hierarchical clustering is a fundamental task often used to discover meaningful structures in data, such as phylogenetic trees, taxonomies of concepts, subtypes of cancer, and cascades of particle decays in particle physics. Typically approximate algorithms are used for inference due to the combinatorial number of possible hierarchical clusterings. In contrast to existing methods, we present novel dynamic-programming algorithms for \emph{exact} inference in hierarchical clustering based on a novel trellis data structure, and we prove that we can exactly compute the partition function, maximum likelihood hierarchy, and marginal probabilities of sub-hierarchies and clusters. Our algorithms scale in time and space proportional to the powerset of $N$ elements which is super-exponentially more efficient than explicitly considering each of the $\exactNumClusterings$ possible hierarchies. Also, for larger datasets where our exact algorithms become infeasible, we introduce an approximate algorithm based on a sparse trellis that compares well to other benchmarks. Exact methods are relevant to data analyses in particle physics and for finding correlations among gene expression in cancer genomics, and we give examples in both areas, where our algorithms outperform greedy and beam search baselines. In addition, we consider Dasgupta's cost \cite{dasgupta2016cost} with synthetic data.
\end{abstract}

% \vspace{-0.5cm}
\blfootnote{Corresponding authors: Craig S. Greenberg , Sebastian Macaluso}
\blfootnote{\href{csgreenberg@cs.umass.edu}{csgreenberg@cs.umass.edu}, \href{sm4511@nyu.edu}{sm4511@nyu.edu}}

\section{Introduction}
Hierarchical clustering is often used to discover meaningful structures, such as phylogenetic trees of organisms \cite{kraskov2005hierarchical}, taxonomies of concepts \cite{cimiano2005learning}, subtypes of cancer \cite{sorlie2001gene},  and jets in particle physics \cite{Cacciari:2008gp}.  
Among the reasons that hierarchical clustering has been found to be broadly useful is that it forms a natural data representation of data generated by a Markov tree, i.e., a tree-shaped model where the state variables are dependent only on their parent or children.
% Additionally, hierarchical clusterings can be used to represent alternative flat partitions of a dataset.

\begin{wrapfigure}{r}{0.4\textwidth} 
\vspace{-20pt}
  \begin{center}
    \includegraphics[width=0.25\textwidth]{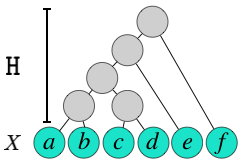}
    \vspace{-0.4cm}
    \caption{\small{Schematic representation of a hierarchical clustering. $\tree$ denotes the latent state and $X$ the dataset. }}
    \label{fig:latentStructure}
  \end{center}
  \vspace{-0.5cm}
%   \vspace{1pt}
\end{wrapfigure}
We define a hierarchical clustering as a recursive splitting of a dataset of $N$ elements, $\dataset = \{x_i\}_{i=1}^N$ into subsets until reaching singletons. This can equivalently be viewed as starting with the set of singletons and repeatedly taking the union of sets until reaching the entire dataset.  We show a schematic representation in Figure \ref{fig:latentStructure}, where we identify each $x_i$ with a leaf of the tree and the latent state as $\tree$. 
% This is in contrast to flat clustering, where the task is to partition the dataset into disjoint subsets.  
% \begin{SCfigure*}
% \vspace{-5pt}
%   \begin{center}
%     \includegraphics[width=0.18\textwidth]{figs/LatentStructure5.png}
%     \vspace{-0.3cm}
%     \caption{\small{Schematic representation of a hierarchical clustering. $\tree$ denotes the latent state and $X$ the dataset. }}
%     \label{fig:latentStructure}
%   \end{center}
%   \vspace{-0.8cm}
% %   \vspace{1pt}
% \end{SCfigure*}

Formally,
\begin{definition}{\textbf{\emph{(Hierarchical Clustering\footnote{We limit our exposition to binary hierarchical clustering. Binary structures encode more tree-consistent clusterings than k-ary \cite{blundell2010bayesian}. Natural extensions may exist for k-ary clustering, which are left for future work.})}}}
  \label{defn:hclustering}
  Given a dataset of elements, $\dataset = \{x_i\}_{i=1}^N$, a
  \textbf{hierarchical clustering}, $\tree$, is a set of nested subsets of $\dataset$, s.t. $\dataset \in \tree$, $\{\{x_i\}\}_{i=1}^N \subset \tree $, and $\forall \dataset_i, \dataset_j \in \tree$, either $\dataset_i \subset \dataset_j$,  $\dataset_j \subset \dataset_i$, or $\dataset_i \bigcap \dataset_j = \emptyset$. Further, $\forall \dataset_i \in \tree$, if  $\exists \dataset_j \in \tree$ s.t. $\dataset_j \subset \dataset_i$, then $\exists \dataset_k \in \tree$ s.t. $\dataset_j \bigcup \dataset_k = \dataset_i$.
\end{definition}

Given a subset $X_L \in \tree$, then $X_L$ is referred to as a cluster in $\tree$. 
When $\dataset_P, \dataset_L, \dataset_R \in \tree$ and $\dataset_L \bigcup \dataset_R = \dataset_P$, we refer to $\dataset_L$ and $\dataset_R$ as children of $\dataset_P$, and $\dataset_P$ the parent of $\dataset_L$ and $\dataset_R$; if $\dataset_L \subset \dataset_P$ we refer to $\dataset_P$ as an ancestor of $\dataset_L$ and $\dataset_L$ a descendent of $\dataset_P$.(We also denote the sibling of $\dataset_L$, as $\dataset_R = \dataset_P \setminus \dataset_L$.)
% We denote $\children(\dataset_P)$ as the children of $\dataset_P$. 
% % In section \ref{sec:trellis} we describe a full and a sparse trellis data structure to do inference in hierarchical clustering. The difference is that the full trellis spans over all possible binary hierarchical clusterings of the dataset $\dataset$ and the sparse one over a subset of them. 
% In general given a parent $X_P$, we define the siblings in $\children(\dataset_P)$ as  $ \textsf{sibs}(X_P) \subseteq \{(X_L,X_R)|X_L \in \children(\dataset_P), X_R \in \children(\dataset_P), X_L \cap X_R = \emptyset, X_L \cup X_R \in X_P\}$. 
For binary trees, the total number of possible pairs of siblings $(X_L,X_R)$ for a parent with $N$ elements is given by the Stirling number of the second kind $S(N,2) = 2^{N-1}-1$. 
% Leaves of the tree refer to individual elements / singleton clusters. 

In our work, we consider an energy-based probabilistic model for hierarchical clustering. 
We provide a general (and flexible) definition of the probabilistic model and then give three specific examples of the distribution in section \ref{experiments}. Our model is based
on measuring the compatibility of all pairs of sibling nodes in a binary tree structure. Formally, 
\begin{definition}
\label{defn:energy_based_hclustering}
\textbf{\emph{(Energy-based Hierarchical Clustering)}}
Let $\dataset$ be a dataset, $\tree$ be a hierarchical clustering of $\dataset$, let $\EfunS: 2^{\dataset} \times 2^{\dataset} \rightarrow \RR^{+}$ be a potential function describing the compatibility of a pair of sibling nodes in $\tree$, and let $\EfunT(\dataset | \tree)$ be a potential function for the $\tree$ structure.
  Then, the probability of $\tree$ for the dataset
   $\dataset$, $P(\tree|\dataset)$, is equal to the unnormalized potential
  of $\tree$ normalized by the partition function,
  $Z(\dataset)$:
%   \vspace{-0.5cm}
  \small{
  \begin{equation}
      P(\tree|\dataset) = \frac{\EfunT(\dataset | \tree)}{Z({\dataset})} \; \text{with} \;
      \EfunT(\dataset | \tree) = \prod_{X_L,X_R \in \textsf{sibs}(\tree)} \EfunS(X_L,X_R)
  \end{equation}
  }
where $ \textsf{sibs}(\tree) = \{(X_L,X_R)|X_L \in \tree, X_R \in \tree, X_L \cap X_R = \emptyset, X_L \cup X_R \in \tree\}$. 
The partition function $Z(\dataset)$ is given by:
\begin{align}
    Z({\dataset}) = \sum_{\tree \in \treeset({\dataset})} {\EfunT(\dataset | \tree)}. \label{eq:Z}
\end{align}
where $\treeset({\dataset})$ represents all binary hierarchical clusterings of the elements $\dataset$. 
\end{definition}

We refer to our model as an energy-based model given that $\EfunS(\cdot, \cdot)$ is often defined by the unnormalized Gibbs distribution, i.e. $\EfunS(X_L, X_R) = \exp(-\beta E(X_L,X_R))$, where $\beta$ is the inverse temperature and $E(\cdot,\cdot)$ is the energy.
This probabilistic model allows us to express many familiar distributions over tree structures. It also has a connection to the classic algorithmic hierarchical clustering technique, agglomerative clustering, in that $\EfunS(\cdot,\cdot)$ has the same signature as a ``linkage function'' (i.e., single, average, complete linkage). We note that in this work we do not use informative prior distributions over trees $P(\tree)$ and instead assume a uniform prior. 

Often, probabilistic approaches, 
such as coalescent models \cite{teh2008bayesian,boyles2012time,hu2013binary} and diffusion trees \cite{neal2003density,knowles2011pitman},
model which tree structures are likely for a given dataset. 
For instance, in particle physics generative models of trees are used to model jets \cite{Cacciari:2008gp}, and similarly coalescent models have been used in phylogenetics \cite{suchard2018bayesian}. Inference in these approaches is done by approximate, rather than exact, methods that lead to local optima, such as greedy best-first, beam-search, sequential Monte Carlo \cite{wang2015bayesian}, and MCMC \cite{neal2003density}. Also, these methods do not have efficient ways to compute an exact normalized distribution over all tree structures.

Exactly performing MAP inference and finding the partition
function by enumerating all hierarchical clusterings over $N$ elements is exceptionally difficult because the number of hierarchies grows extremely rapidly, namely $\exactNumClusterings$ (see \cite{callan2009combinatorial,DaleMoonCatalanSets} for more details and proof), where $!!$ is double factorial. 
To overcome the computational burden, in this paper we introduce a cluster trellis data structure for hierarchical clustering. The cluster trellis, inspired by \cite{NIPS2018_8081}, 
enables us to use dynamic programming algorithms
to exactly
compute MAP structures and the partition function, as well as compute marginal distributions, including the probability of any sub-hierarchy or cluster.
We further show how to sample exactly from the posterior distribution over hierarchical clusterings (i.e., the probability of sampling a given hierarchy is equal to the probability of that hierarchy).
Our algorithms compute these quantities
without having to iterate over each possible hierarchy
in the $\mathcal{O}(3^{N})$ time, which is super-exponentially more efficient than explicitly considering each of the $\exactNumClusterings$ possible hierarchies (see Corollary \ref{coroll:improvement} for more details). 
% While still exponential, this is orders of magnitude faster than enumerating all 
% trees and is to our knowledge the fastest exact MAP / partition function result (See \S \ref{proof:map-complexity} and \S \ref{proof:Z-time-complexity} for proofs).
Thus, while still exponential, this is  feasible in regimes where enumerating all possible trees would be infeasible, and is to our knowledge the fastest exact MAP/partition function result(See \S \ref{proof:map-complexity} and \S \ref{proof:Z-time-complexity} for proofs), making practical exact inference for datasets on the order of 20 points ($\sim 3\times10^9$ operations vs $\sim 10^{22}$ trees) or fewer.
For larger datasets, we introduce an approximate algorithm based on a sparse hierarchical cluster trellis and we outline different strategies for building this sparse trellis.
We demonstrate our methods' capabilities for exact inference in discovering cascades of particle decays in jet physics and subtype hierarchies in cancer genomics, two applications where there is a need for exact inference on datasets made feasible by our methods.
We find that greedy and beam search methods frequently return estimates that are sub-optimal compared to the exact MAP clustering.
\vspace{-0.2cm}
\paragraph{Contributions of this Paper.}
We achieve \emph{exact}, not approximate, solutions to the following:
\begin{itemize}
\setlength{\itemsep}{-1pt}%
    \vspace{-0.3cm}
    \item \textbf{Compute the Partition Function} $Z(\dataset)$.
    \item \textbf{MAP Inference}, i.e. find the maximum likelihood tree structure $\argmax_{\tree \in \treeset} P(\tree|\dataset)$.
    \item \textbf{Sample Hierarchies from the Posterior Distribution}, i.e. weighted by their probability, $P(\tree|\dataset)$.
\end{itemize}
\vspace{-0.2cm}
\vspace{-0.1cm}
\section{Hierarchical Cluster Trellis}\label{sec:trellis}
\vspace{-0.2cm}
% Exactly performing MAP inference and finding the partition
% function by enumerating all hierarchical clusterings over $N$ elements is exceptionally difficult because the number of hierarchies grows extremely rapidly, namely $\exactNumClusterings$ (see \cite{callan2009combinatorial,DaleMoonCatalanSets} for more details and proof), where $!!$ is double factorial. 
% To overcome the computational burden, we introduce a cluster trellis data structure \cite{NIPS2018_8081} for hierarchical clustering. We describe how this data structure
% enables us to use dynamic programming algorithms
% to exactly
% compute MAP structures and the partition function.
% Our algorithms compute these quantities
% without having to iterate over each possible hierarchy
% in the $\mathcal{O}(3^{N})$ time. While still exponential, this is orders of magnitude faster than enumerating all 
% trees and is to our knowledge the fastest exact MAP / partition function result (See \S \ref{proof:map-complexity} and \S \ref{proof:Z-time-complexity} for proofs).
% Furthermore, we demonstrate how our methods can be used to sample structures from $P(\tree|\dataset)$, compute marginal probabilities of subtrees, all without enumerating the complete set of hierarchical clusterings.  

\subsection{Trellis Data Structure}
\vspace{-0.2cm}
The trellis data structure is a directed acyclic graph.
% , where there is a bijection between the vertices of the graph and $\powerset(\dataset)$, and there is an edge from vertex $\trellisVertex_i$ to vertex $\trellisVertex_j$ when the set of elements associated with $\trellisVertex_j$ is a maximal superset of the set of elements associated with $\trellisVertex_i$. 
The dataset associated with a trellis vertex $\trellisVertex$ is denoted $\dataset(\trellisVertex)$ and the trellis vertex associated with a dataset $\dataset$ is denoted $\trellisVertex(\dataset)$. Each vertex in the trellis will store memoized values of $Z(\trellisVertex)$ for computing the partition function, as well as the value $\EfunT(\tree^*[\trellisVertex])$ and the backpointer $\Xi(\tree^*[\trellisVertex])$ for computing the MAP tree. 
We denote $\children(\dataset)$ as the children of $\trellisVertex(\dataset)$. 
% We note that the edges of the structure are not crucial for presentation here. However they are practically useful in the implementation of the trellis and provide a parallel between the trellis structure and a DAG structure and the tree structure given by a hierarchical clustering. 
We refer to a full trellis as the data structure where every possible hierarchical clustering given a dataset $\dataset$ can be realised. In contrast, a sparse trellis will only contain a subset of all possible hierarchies.
\vspace{-0.2cm}
\subsection{Computing the Partition Function}
\label{sec:partition_function}
\vspace{-0.2cm}
Given a dataset of elements, $\dataset = \{x_i\}_{i=1}^N$, the partition function, $Z({\dataset})$, for the set of hierarchical clusterings over $\dataset$, $\treeset(\dataset)$, is given by Equation \ref{eq:Z}. 
The trellis implements a memoized 
dynamic program to compute the partition function
and the MAP. To achieve this, we need to re-write the 
partition function in the corresponding recursive
way. In particular, 
\vspace{-0.05cm}
\begin{proposition}
        \label{thm:recursive-partition}
        For any $x \in \dataset$, the hierarchical partition function can be written recursively, as
        $
        Z(\dataset) =  \sum_{\tree \in \treeset(\dataset)}{\EfunT(\dataset | \tree)} = {\sum_{\dataset_i \in  \children(\dataset)_x}}{\EfunS(\dataset_i, \dataset \setminus \dataset_i) \cdot Z(\dataset_i) \cdot Z(\dataset \setminus \dataset_i)}
       $
        where $\children(\dataset)_x$ is the set of all children of $\dataset$ containing the element $x$, 
       i.e,. $\children(\dataset)_x = \{\dataset_j : \dataset_j \in \children(\dataset)  \land x \in \dataset_j \}$. In the particular case of a full trellis, then $\children(\dataset)_x = \{\dataset_j : \dataset_j \in 2^{\dataset}\setminus\dataset \land x \in \dataset_j \}$.
\end{proposition}
% \begin{proposition}
%         \label{thm:recursive-partition}
%         For any $x \in \dataset$, the hierarchical partition function can be written recursively, as
%         $
%         Z(\dataset) =  \sum_{\tree \in \treeset(\dataset)}{\EfunT(\dataset | \tree)} = {\sum_{\dataset_i \in  \dataset_x}}{\EfunS(\dataset_i, \dataset \setminus \dataset_i) \cdot Z(\dataset_i) \cdot Z(\dataset \setminus \dataset_i)}
%       $
%         where $\dataset_x$ is the set of all clusters containing the element $x$ (omitting \dataset), 
%       i.e,. $\dataset_x = \{\dataset_j : \dataset_j \in 2^{\dataset}\setminus\dataset \land x \in \dataset_j \}$.
% \end{proposition}
\begin{figure*}
\vspace{-0.2cm}
        \centering
        \includegraphics[width=1\textwidth]{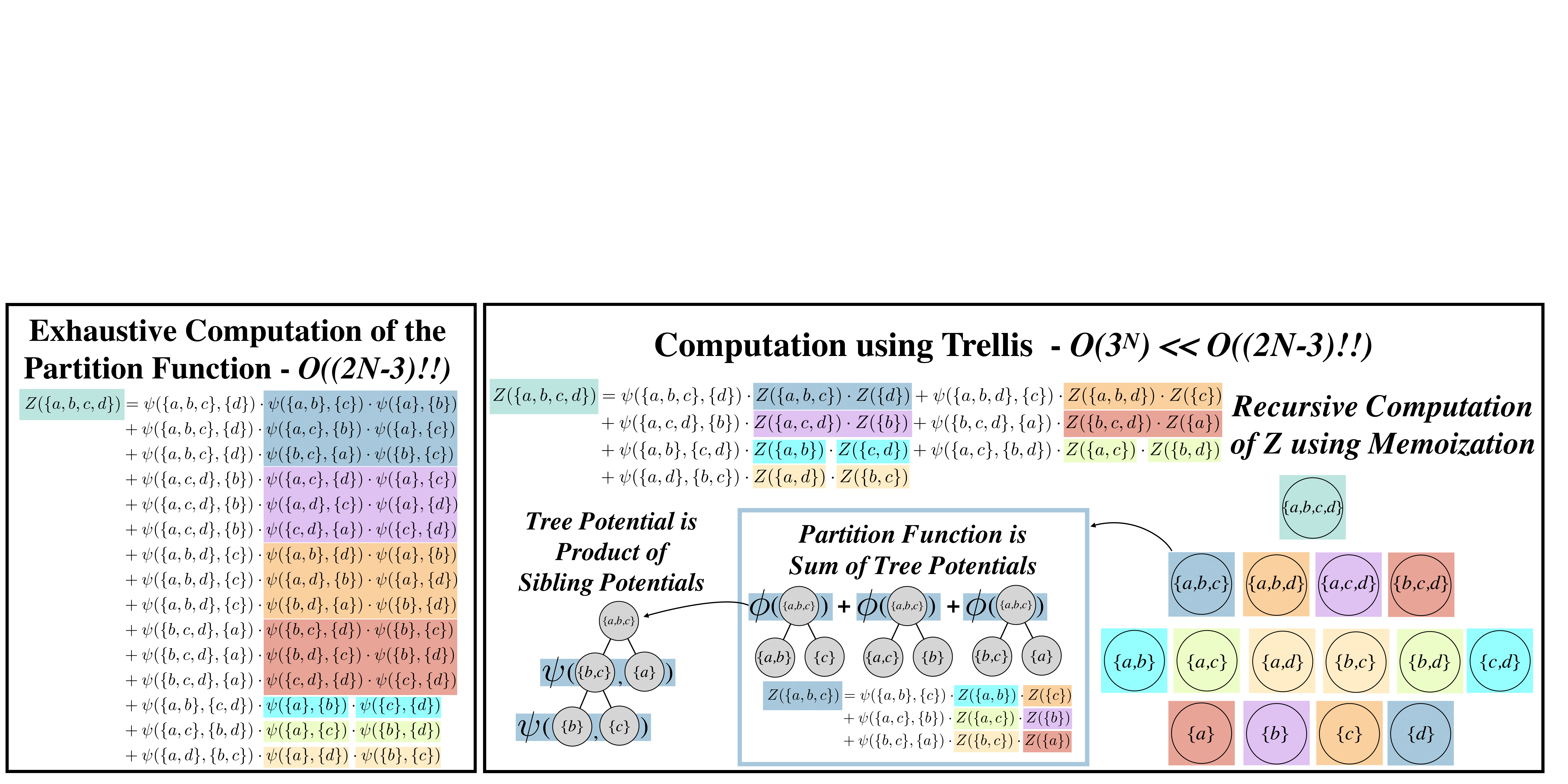} 
    % \caption{\textbf{Computing the partition function}. An example of using a trellis to compute the distribution over hierarchical clusterings of the dataset $\{a,b,c,d\}$.  The left panel shows the exhaustive computation of the partition function, consisting of the summation of $(2\cdot4-3)!!$ energy equations, one for each of the $5!!=15$ trees rooted at $\{a,b,c,d\}$. The right panel shows the computation of the partition function using the corresponding trellis.  The sum for the partition function is over $2^{4-1} - 1 = 7$ equations, each making use of a memoized Z value. Colors indicate corresponding computations that are computed with and stored in the trellis.}
        \caption{\small{\textbf{Computing the partition function for the dataset $\{a,b,c,d\}$}. Left: exhaustive computation, consisting of the summation of $(2\cdot4-3)!! =15 $ energy equations. Right: computation using the trellis.  The sum for the partition function is over $2^{4-1} - 1 = 7$ equations, each making use of a memoized $Z$ value. Colors indicate corresponding computations over siblings in the trellis.}}
\label{fig:trellis_example_b}
\vspace{-0.45cm}
\end{figure*}
The proof is given in $\S$ \ref{thm:recursive-partition-proof} in the Appendix. 
Algorithm \ref{alg:partition_function_trellis} 
describes in a recursive way how to efficiently compute the partition function using the trellis based on Proposition \ref{thm:recursive-partition}. We first set the partition function of the leaf nodes in the trellis to 1. 
% Algorithm \ref{alg:partition_function_trellis} is described in a recursive way. 
% Then, we start by picking any element in the dataset, $x_i$, and the complement containing all points other than $x_i$, $\dataset \setminus x_i$. It then considers all clusters, $\dataset_j$, of the powerset of $X \setminus x_i$, i.e., $\dataset_j \in 2^{\dataset_j \setminus x_i})$. 
% For the cluster $X_i = X_j \cup \{x_i\}$, the partition function is computed (memoized, recursively) for $X_i$ and its complement, thus enabling the application of Proposition \ref{thm:recursive-partition} to get $Z(X)$.
Then, we start by selecting any element in the dataset, $x_i$, and consider all clusters $\dataset_i \in \children(\dataset)$ such that $x_i \in \dataset_i$. Next, the partition function is computed (memoized, recursively) for $X_i$ and its complement $\dataset \setminus \dataset_i$, thus enabling the application of Proposition \ref{thm:recursive-partition} to get $Z(X)$.
For a full trellis, the algorithm can straightforwardly be written in a bottom-up, non-recursive way. 
In this case, the partition function for every node in the trellis is computed in order (in a bottom-up approach), from the nodes with the smallest number of elements to the nodes with the largest number of elements, memoizing the partition function value at each node.  By computing the partial partition functions in this order, whenever computing the partition function of a given node in the trellis, the corresponding ones of all of the descendent nodes will have already been computed and memoized. 
In Figure \ref{fig:trellis_example_b}, we show a visualization comparing the computation of the partition function with the trellis to the brute force method for a dataset of four elements. Next, we present the complexity result for Algorithm \ref{alg:partition_function_trellis}:
% \vspace{-0.6cm}
\begin{algorithm}
\caption{\texttt{PartitionFunction}$(\dataset)$}
\label{alg:partition_function_trellis}
\begin{algorithmic}
    \STATE Pick $x_i \in \dataset$ \text{ and set }  $Z(\dataset) \gets 0$
    \FOR{ $\dataset_i$ in $\children(\dataset)_{x_i}$} 
    % \FOR{$$\dataset_j$ in $2^{\dataset \setminus \{x_i\}}$}
    % \STATE $\dataset_i \gets \dataset_j \bigcup \ \{x_i\}$
    \STATE \textbf{if} {$Z(\dataset_i)$ not set}  \textbf{then} \\$Z(\dataset_i) \gets \texttt{PartitionFunction}(\dataset_i)$
    \STATE{\textbf{if} $Z(\dataset \setminus \dataset_i)$ not set} \textbf{then} \\ $Z(\dataset \setminus \dataset_i) \gets \texttt{PartitionFunction}(\dataset \setminus \dataset_i)$
    \STATE \small{$Z(\dataset) \leftarrow Z(\dataset) + \EfunS(\dataset_i, \dataset \setminus \dataset_i) \cdot Z(\dataset_i) \cdot  Z(\dataset \setminus \dataset_i)$}
    \ENDFOR
    \STATE \textbf{return} $Z(\dataset)$
\end{algorithmic}
\end{algorithm}
% \begin{algorithm}[tb]
% \caption{\texttt{PartitionFunction}$(\dataset)$}
% \label{alg:partition_function_trellis}
% \begin{algorithmic}
%     \STATE Pick $x_i \in \dataset$ \text{ and set }  $Z(\dataset) \gets 0$
%     \FOR{\sebastian{ $\dataset_i$ in $\children(\dataset)_{x_i}$} }
%     % \FOR{$$\dataset_j$ in $2^{\dataset \setminus \{x_i\}}$}
%     % \STATE $\dataset_i \gets \dataset_j \bigcup \ \{x_i\}$
%     \STATE \textbf{if} {$Z(\trellisVertex(\dataset_i))$ not set}  \textbf{then} \\$Z(\trellisVertex(\dataset_i)) \gets \texttt{PartitionFunction}(\dataset_i)$
%     \STATE{\textbf{if} $Z(\trellisVertex(\dataset \setminus \dataset_i))$ not set} \textbf{then} \\ $Z(\trellisVertex(\dataset \setminus \dataset_i) \gets \texttt{PartitionFunction}(\dataset \setminus \dataset_i)$
%     \STATE \small{$Z(\dataset) \leftarrow Z(\dataset) + \EfunS(\dataset_i, \dataset \setminus \dataset_i) \cdot Z(\trellisVertex(\dataset_i)) \cdot  Z(\trellisVertex(\dataset \setminus \dataset_i))$}
%     \ENDFOR
%     \STATE \textbf{return} $Z(\dataset)$
% \end{algorithmic}
% \end{algorithm}
\begin{theorem}
\label{thm:partition-function-time-complexity}
For a given dataset $\dataset$ of $N$ elements, Algorithm \ref{alg:partition_function_trellis} computes $Z(\dataset)$ in $\mathcal{O}(3^N)$ time.
\end{theorem}
The time-complexity of the algorithm is $\mathcal{O}(3^{N})$, which is is significantly smaller than the $\exactNumClusterings$ possible hierarchies.  
\begin{corollary}
\label{coroll:improvement}
For a given dataset $\dataset$ of $N$ elements, Algorithm \ref{alg:partition_function_trellis} is super-exponentially more efficient than brute force methods that consider every possible hierarchy. In particular the ratio is $\mathcal{O}((\frac{2}{3})^N \,\,\Gamma(N - 1/2))$.
\end{corollary}
\vspace{-0.15cm}
The proofs of Algorithm \ref{thm:partition-function-time-complexity} and Corollary \ref{coroll:improvement} are given in $\S$ \ref{proof:Z-time-complexity} of the Appendix.
% \sebastian{: remove the following as doesn't explain the complexity? "since the partition function for each node in the trellis is computed and the descendent nodes' partition functions are always pre-computed due to the order of computation. "}
\vspace{-0.2cm}
\subsection{Computing the MAP Hierarchical Clustering}
\vspace{-0.2cm}
Similar to other dynamic programming algorithms, such as Viterbi, we can adapt Algorithm \ref{alg:partition_function_trellis} in order to find the MAP hierarchical clustering.
The MAP clustering for dataset $\dataset$, is $\treestar(\dataset) = \argmax_{\tree \in \treeset(\dataset)}{\EfunT(\tree)}$.
Here we can also use a recursive memoized technique, where each node will store a value for the MAP, denoted by $\EfunT(\treestar(\dataset))$ and a backpointer $\Xi(\treestar(\dataset))$. Specifically,
% To use the recursive technique we use the following Proposition for correctness of the recursion.
\begin{proposition}
        \label{thm:recursive-map}
        For any $x \in \children(\dataset)$, let $\children(\dataset)_x = \{\dataset_j : \dataset_j \in \children(\dataset)  \land x \in \dataset_j \}$, then 
       $ \EfunT(\treestar(\dataset))  =  \max_{X_i \in \children(X)_x} \EfunS(\dataset_i, \dataset \setminus \dataset_i) \cdot \EfunT(\treestar(X_i)) \cdot \EfunT(\treestar(\datasetcomplement{X_i})) $.
\end{proposition}
% \begin{proposition}
%         \label{thm:recursive-map}
%         For any $x \in \children(\dataset)$, let $X_x = \{X_j : X_j \in 2^\dataset \setminus X \land x \in X_j \}$, then 
%       $ \EfunT(\treestar(\dataset))  =  \max_{X_i \in X_x} \EfunS(\dataset_i, \dataset \setminus \dataset_i) \cdot \EfunT(\treestar(X_i)) \cdot \EfunT(\treestar(\datasetcomplement{X_i})) $.
% \end{proposition}
\vspace{-0.15cm}
See \S \ref{proof:recursive-map} in the Appendix for the proof. As in the partition function algorithm described in Section \ref{sec:partition_function}, the time complexity for finding the MAP clustering is also $\mathcal{O}(3^{N})$. The main difference is that to compute the maximal likelihood hierarchical clustering, the maximal energy of the sub-hierarchy rooted at each node is computed, instead of the partition function.  Pointers to the children of the maximal sub-hierarchy rooted at each node are stored at that node.  A proof of the time complexity, analogous to the one for the partition function, can be found in \S \ref{proof:map-complexity} of the Appendix.
\vspace{-0.2cm}
\subsection{Computing Marginals}
\vspace{-0.2cm}
In this section, we describe how to compute two types of marginal probabilities. The first is for a given sub-hierarchy rooted at $\dataset_i$, i.e.,
$\tree_i \in \treeset(\dataset_i)$, defined as ${P(\tree_i|\dataset) = {\sum_{\tree \in A(\tree_i)}} P(\tree|\dataset) }$, where $A(\tree_i) = \{\tree : \tree \in \treeset(\dataset) \land \tree_i \subset \tree\}$, and $\tree_i \subset \tree$ indicates that $\tree_i$ is a subtree of $\tree$. 
Thus, we marginalize over every possible hierarchy while keeping fixed the sub-hierarchy $\tree_i$. The second is for a given cluster, $\dataset_i$, defined as 
${P(\dataset_i|\dataset) = {\sum_{\tree \in A(\dataset_i)}} P(\tree|\dataset) }$, where $A(\dataset_i) = \{\tree : \tree \in \treeset(\dataset) \land \dataset_i \subset \tree\}$, and $\dataset_i \subset \tree$ indicates that cluster $\dataset_i$ is contained in $\tree$. In this case, we marginalize over every possible sub-hierarchy that contains the cluster $\dataset_i$ while keeping the rest of the hierarchy $\tree$ fixed.
% $P(\dataset_i|\dataset) = {\sum_{\tree_i \in \treeset(\dataset_i)}}{P(\tree_i|\dataset_i)}$.
The value of $P(\tree_i|X)$ can be computed using the same algorithm used for the partition function, except that we first merge $\tree_i$ into a single leaf node and use $\EfunT(\tree_i(\dataset_i))$ for the energy of the newly merged leaf. 
The same is true for computing the value of $P(\dataset_i|\dataset)$, except that after merging $\dataset_i$ into a single leaf node, the value $Z(\dataset_i)$ should be used. 
See Appendix $\S$ \ref{proof:marginal-formulas} for proofs.
% The second is for
% a given cluster, $\dataset_i$, defined as 
% $P(\dataset_i|\dataset) = {\sum_{\tree_i \in \treeset(\dataset_i)}}{P(\tree_i|\dataset)}$.
% The value of $P(\tree_i|X_i)$ can be computed using the same algorithm used for the partition function, except by first merging $\dataset(\tree_i|\dataset)$ into a single leaf node, but using $\EfunT(\dataset(\tree_i))$ for the energy of the newly merged leaf. 
% The same is true for computing the value of $P(\dataset_i|\dataset)$, except after merging $\dataset_i$ into a single leaf node, the value $Z(\dataset_i)$ should be used. 
% See Appendix $\S$ \ref{proof:marginal-formulas} for proofs.
\vspace{-0.2cm}
\subsection{Sampling from the Posterior Distribution}\label{sec:Treesampling}
\vspace{-0.2cm}
Drawing samples from the true posterior distribution 
$P(\tree|\dataset)$ is also difficult because of the 
extremely large number of trees. In this section, 
we introduce a sampling procedure for hierarchical clusterings $\tree_i$ implemented using the trellis which gives samples from the exact true posterior without enumerating all possible hierarchies.

The sampling procedure will build a tree structure in a top-down way. We start with the cluster of all the elements, $\dataset$, then sample one child of that cluster, $\dataset_L \subset \dataset$, (Eq. \ref{eq:levelSampling}) and set the other one to be the complement of $\dataset_L$, i.e., $\dataset \setminus \dataset_L$. This is repeated recursively from each of the children and terminates when a cluster contains a single element. A child  $\dataset_L$ of parent $X_p$, i.e., $X_L \subset X_p$ is sampled according to:
\vspace{-0.2cm}
\begin{equation}\label{eq:levelSampling}
\small{p(\dataset_L| \dataset_p) = \frac{1}{Z(\dataset_p)} \cdot \, \EfunS(X_L,\dataset_p \backslash X_L)\, \cdot  Z(\dataset_L) \cdot Z(\dataset_p \backslash \dataset_L).}
\end{equation}
\vspace{-0.1cm}
Pseudocode for this algorithm is given in Algorithm \ref{alg:sampling_function_trellis}. 
\begin{theorem}
\label{thm:sampling}
\emph{ \texttt{Sample}}$(\dataset)$ (Alg. \ref{alg:sampling_function_trellis}) gives samples from $P(\tree|\dataset)$.
\end{theorem}
\vspace{-0.15cm}
The proof is given in Appendix $\S$ \ref{proof:sampling-method}. This algorithm is notable in that it does not require computing a categorical distribution over all trees and samples exactly according to $P(\tree|\dataset)$.
\vspace{-0.25cm}

\begin{algorithm}[tb]
  \caption{\texttt{Sample}$(\dataset)$}
  \label{alg:sampling_function_trellis}
\begin{algorithmic}
    \STATE \textbf{if} {$ |X| = 1$ } \textbf{return} $\{X\}$
    \STATE Sample $X_L$ from  $p(X_i | X)$ (Eq. \ref{eq:levelSampling}).
    \STATE \textbf{return} $\{X_L,\ X \setminus X_L\} \ \cup \ \texttt{Sample}(X_L)\ \cup \ \texttt{Sample}(X \setminus X_L)$ 
\end{algorithmic}
\end{algorithm}

\vspace{-0.2cm}

\section{Sparse Hierarchical Cluster Trellis}\label{sec:sparsetrellis}
\vspace{-0.2cm}
%  While the trellis is efficient compared to the factorial growth in the number of possible hierarchical clusterings, the scaling of this algorithm is limited to small datasets. In this section we introduce an approximate algorithm based on a sparse hierarchical trellis that can scale to larger datasets and we outline different strategies for building this sparse trellis.
In this section, we introduce a sparse trellis data structure, which allows to scale to larger datasets by controlling the sparsity index, i.e. the fraction of hierarchies we consider from the total of $\exactNumClusterings$. Most hierarchies have potential values orders of magnitude smaller than the MAP clustering making their contribution to the partition function negligible. 
%  \footnote{In general, low likelihood is equivalent to large energies. However, we refer to likelihood as this is directly connected to the jet physics example used as a case study in this paper.}. 
As a result, if we build a sparse trellis that considers the most relevant hierarchies, we could find approximate solutions for inference
% the MAP values and partition function for 
in datasets where implementing the full trellis is not feasible.
Conceptually, the only difference with respect to the full trellis is that the children of each vertex are typically a subset of all $2^\dataset$ possible ones. Thus, the algorithms and proofs are the same as the ones presented in section \ref{sec:trellis} but the solutions will be approximate. The specific vertices that are contained in the sparse trellis depend on how we build it. Below we present two possible strategies.
% The algorithms are similar to the exact cluster trellis ones, though there are differences in the implementation.
\vspace{-0.3cm}
\subsection{Building Strategies}\label{sec:building}
\vspace{-0.2cm}
The performance of the sparse trellis depends on the subset of all possible hierarchies over which it expands. This subset is chosen by a building strategy, which provides a sample of trees used to create the trellis. 
There are also different mappings for the ordering of the leaves of the input trees, and it is interesting to study the different subsets of hierarchies spanned by the sparse trellis depending on this mapping.
% For this paper, we ordered the leaves in increasing norm of their momentum vector $\vec{p} \in \RR^3$ (see section \ref{Uncertainty in Clustering} for more details about the model).

\begin{figure}
\vspace{-0.3cm}
\centering
\includegraphics[width=0.6\columnwidth]{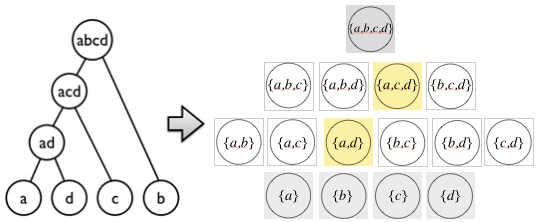}
\vspace{-0.2cm}
\caption{\small{Schematic representation of how the sparse trellis is built iterating over each tree with four leaves from a sample dataset $\dataset$. After every hierarchical structure is added, the final trellis is composed of the colored vertices, the added edges, the leaves and the root vertex. The vertices that are not colored represent the subset of vertices of the full trellis that are missing in the sparse case.
}}
\label{fig:sparseTrellis}
\vspace{-0.3cm}
\end{figure}

We start with a set of input trees.
Once we choose a specific ordering of the leaves, we iterate over each input tree, creating a vertex $\trellisVertex_i$ in the trellis for each new node in the tree, i.e. nodes that have not been visited in previous input trees. A schematic representation is shown in Figure \ref{fig:sparseTrellis}. This way, the input sample of trees determines the trellis vertices that are created. The trellis considers every possible hierarchical clustering that can be realized with these vertices which is typically much greater than the number of input trees.  
After creating the trellis, we initialize the leaf vertices values with some dataset of interest and run the inference algorithms, e.g. MAP and partition function computations.

Next, we present two distinctive procedures to build the trellis, and name the resulting trellises.
%according to them as Simulator and Beam Search trellis.

{\bf Simulator Trellis:} In some cases there exists a generative model or simulator that implicitly defines a distribution over hierarchies. We can use trees sampled from this simulator to seed the sparse trellis.
% In this case, sample trees are expected to be around the mean of the posterior distribution. 
We restrict the generated trees to have the same number of leaves, which is fixed for a given sparse trellis.
%each trellis we create.

{\bf Beam Search Trellis:} input trees are obtained from running the beam search algorithm over a sample of sets of leaves. This approach is much more general, as it could be implemented for datasets where there is no generative model. 
% In this case, we expect the distribution for the likelihood of our input trees to be shifted toward higher values, compared to the simulator trellis.
Note that we choose beam search for our experiments, but this approach could be implemented with any  agglomerative clustering, and only requires a “linkage function” (i.e., single, average, complete linkage).
\vspace{-0.2cm}
\section{Experiments}\label{experiments}
\vspace{-0.2cm}
In this section, we demonstrate the use of the exact MAP, partition function, and sampling approaches described in this paper on two real world applications: jet physics and cancer genomics, as well as one synthetic data experiment related to Dasgupta's cost \cite{dasgupta2016cost}. First, we give an illustrative example for the use of the proposed approaches with Dasgupta's cost, running on the kinds of data for which greedy methods are known to be approximate. In each real world application, we demonstrate how the trellis is used to compute exact MAP and the distribution over clusterings that are more informative and accurate than approximate methods. In particle physics, we 
% use a simulation model for cascades of particle physics decays in jet physics that provides ground truth hierarchies facilitating evaluation. We 
additionally demonstrate the use of the sampling procedure (\S \ref{sec:Treesampling}) and the implementation of a sparse trellis. In cancer genomics, we show how we can model subtypes of cancer, which can help determine prognosis and treatment plans. 

\vspace{-0.2cm}
\subsection{Dasgupta's Cost}
\begin{SCfigure*}
% \vspace{-0.2cm}
    \centering
    \caption{{\small \textbf{Dasgupta's Cost}. Trellis vs agglomerative clustering MAP trees for a graph that is known to be difficult for with greedy methods.}}
            \includegraphics[width=0.7\textwidth]{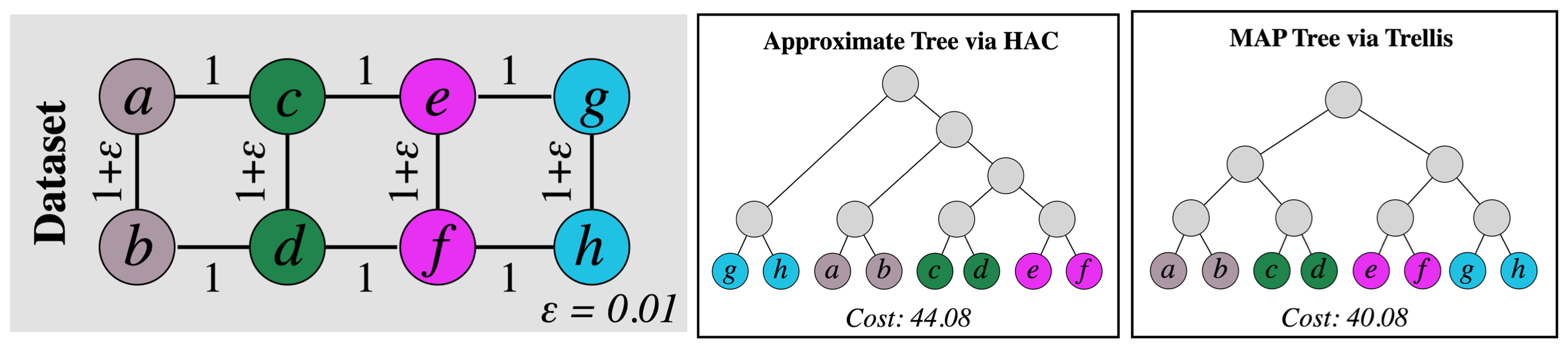} 
% \vspace{0.2cm}
    \label{fig:dasgupta-cost-trees}
    % \vspace{-0.2cm}
\end{SCfigure*}

\vspace{-0.2cm}
\paragraph{Probabilistic model}
\cite{dasgupta2016cost} defines a cost function for hierarchical clustering that has been the subject of much theoretical interest (primarily on approximation algorithms for the cost)   \cite{,cohen2017hierarchical,cohen2019hierarchical,charikar2017approximate,charikar2019hierarchical,moseley2017approximation,roy2017hierarchical}.
Given a graph with vertices of the dataset $\dataset$ and weighted edges representing pairwise similarities between points $\mathcal{W} = \{(i,j,w_{ij}) | i,j \in \{1,..., |\dataset|\} \times \{1,..., |\dataset|\}, i< j, w_{ij} \in \RR^{+}\}$. Dasgupta's cost is defined as:
\begin{equation}
\vspace{-0.1cm}
E(X_i,X_j) = (|X_i| + |X_j|) \sum_{x_i, x_j \in X_i \times X_j} w_{ij} 
\label{eq:dasgupta_potential}
\end{equation}
\vspace{-0.1cm}
This is equivalent to the cut-cost definition of Dasgupta's cost with the restriction to binary trees \cite{dasgupta2016cost}. 
% \vspace{-0.1cm}
\paragraph{Results} Figure \ref{fig:dasgupta-cost-trees} gives an example graph, as proposed by \cite{charikar2019hierarchical} to bound average-linkage performance, following a model for which greedy methods are known to be approximate with respect to Dasgupta's cost  \cite{moseley2017approximation,cohen2017hierarchical}. We run greedy agglomerative clustering and trellis-based MAP procedure (Eq. \ref{eq:dasgupta_potential}). Unsurprisingly, the greedy method fails to achieve the lowest cost tree while the trellis-based method identifies an optimal tree. The cost of the greedily built tree is 44.08 while the tree built using the trellis is 40.08.

\vspace{-0.1cm}
\subsection{Jet Physics} \label{exp:jets}
\vspace{-0.1cm}
\paragraph{Background} The Large Hadron Collider (LHC) at CERN collides two beams of high-energy protons and produces many new  (unstable) particles. Some of these new particles (quarks and gluons) will undergo a {\it showering process}, where they radiate many other quarks and gluons in successive binary splittings. These $1 \rightarrow 2$ splittings can be represented with a binary tree, where the energy of the particles decreases after each step. When the energy is below a given threshold, the showering terminates, resulting in a spray of particles that is called a \textit{jet}. The particle detectors only observe the leaves of this binary tree (the jet constituents), and the unstable particles in the showering process are unobserved. Thus, a  specific jet could result from several latent trees generated by the showering process. While the latent showering process is unobserved, it is described by quantum chromodynamics (QCD).
\vspace{-0.1cm}
\paragraph{Probabilistic Model}
The potential of a hierarchy is identified with the product of the likelihoods of all the  $1\rightarrow 2$ splittings of a parent cluster into two child clusters in the binary tree. 
Each cluster, $\dataset$, corresponds to a particle with an energy-momentum vector $x = (E \in \RR^{+}, \vec{p} \in \RR^3)$ and squared mass ${t}(x) = {E^2-|\vec{p}|^2}$. 
A parent's energy-momentum vector is obtained from adding its children, i.e., $x_P = x_L + x_R$.
We study a toy model for jet physics \cite{ToyJetsShower}, where for each pair of parent and left (right) child cluster with masses $\sqrt{t_P}$ and $\sqrt{t_L}$ ($\sqrt{t_R}$) respectively, the likelihood function is,
    \begin{align}
        \EfunS(X_L, X_R)= f(t(x_L) | t_P, \lambda) \cdot f(t(x_R) | t_P, \lambda)\\ \text{with} \,\,\,\,\,\,
    % \end{align}
    % where
    % \begin{equation}
        f(t | t_P, \lambda) = \frac{1}{1 - e^{- \lambda}} \frac{\lambda}{t_P} e^{- \lambda \frac{t}{t_P}}
    % \end{equation}
    \end{align}
    where the first term in $f(t | t_P, \lambda)$ is a normalization factor associated to the constraint that $t<t_P$. 
    % Thus, the likelihood of a $1\rightarrow 2$ splitting is 
\vspace{-0.1cm}
\paragraph{Data and Methods}
% In this paper, we proposed a new method to efficiently find the MAP hierarchical clustering, partition function Z, and sample hierarchies from the exact true posterior distribution. 
We will compare full and sparse trellises results for the MAP hierarchical clustering with approximate methods, as described below.
The ground truth hierarchical clusterings of our dataset are generated with the toy generative model for jets Ginkgo, see \cite{ToyJetsShower} for more details. This model implements a recursive algorithm to generate a binary tree, whose leaves are the jet constituents. Jet constituents (leaves) and intermediate state particles (inner nodes) in Ginkgo are represented by a four dimensional energy-momentum vector.  

\begin{figure}
    \centering
    \begin{minipage}{0.45\textwidth}
        \centering
        \includegraphics[width=0.99\columnwidth]{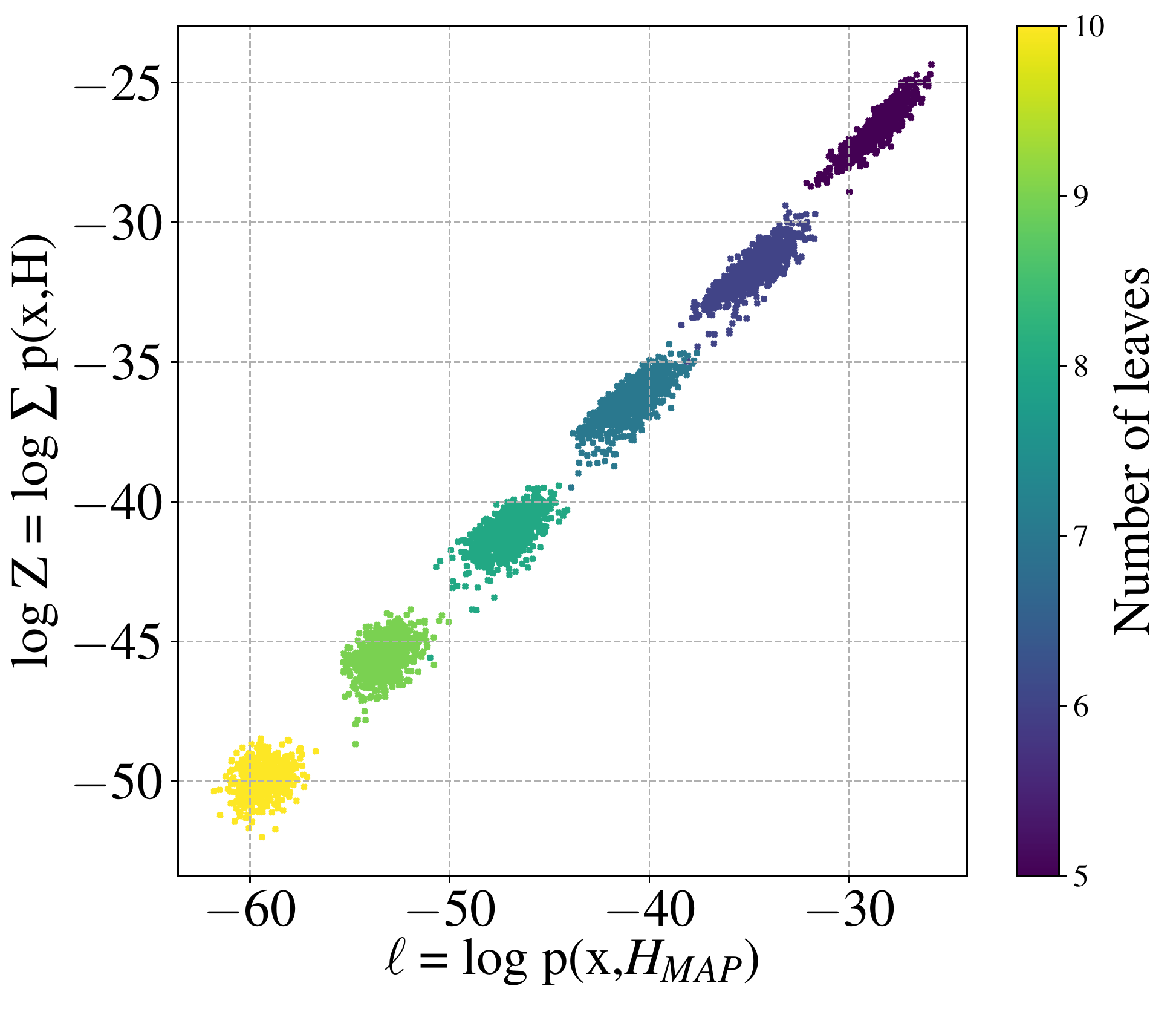}
        \caption{\small{Scatter plot of the partition function $Z$  vs. the trellis MAP value $\ell$ for Ginkgo510 dataset, with up to 10 leaves (jet constituents). The color indicates the number of leaves of each hierarchical clustering. There appears to be a correlation between $Z$ and the MAP values.
        % that is dependent on the number of leaves. 
        % Also, the greater the number of leaves, the faster that $Z$ grows which is consistent with the rapid increase in the number of possible hierarchical clusterings.
        }}
        \label{fig:ZvsMLEscatter}
    \end{minipage}\hfill
    \begin{minipage}{0.45\textwidth}
        \centering
        \includegraphics[width=0.99\columnwidth]{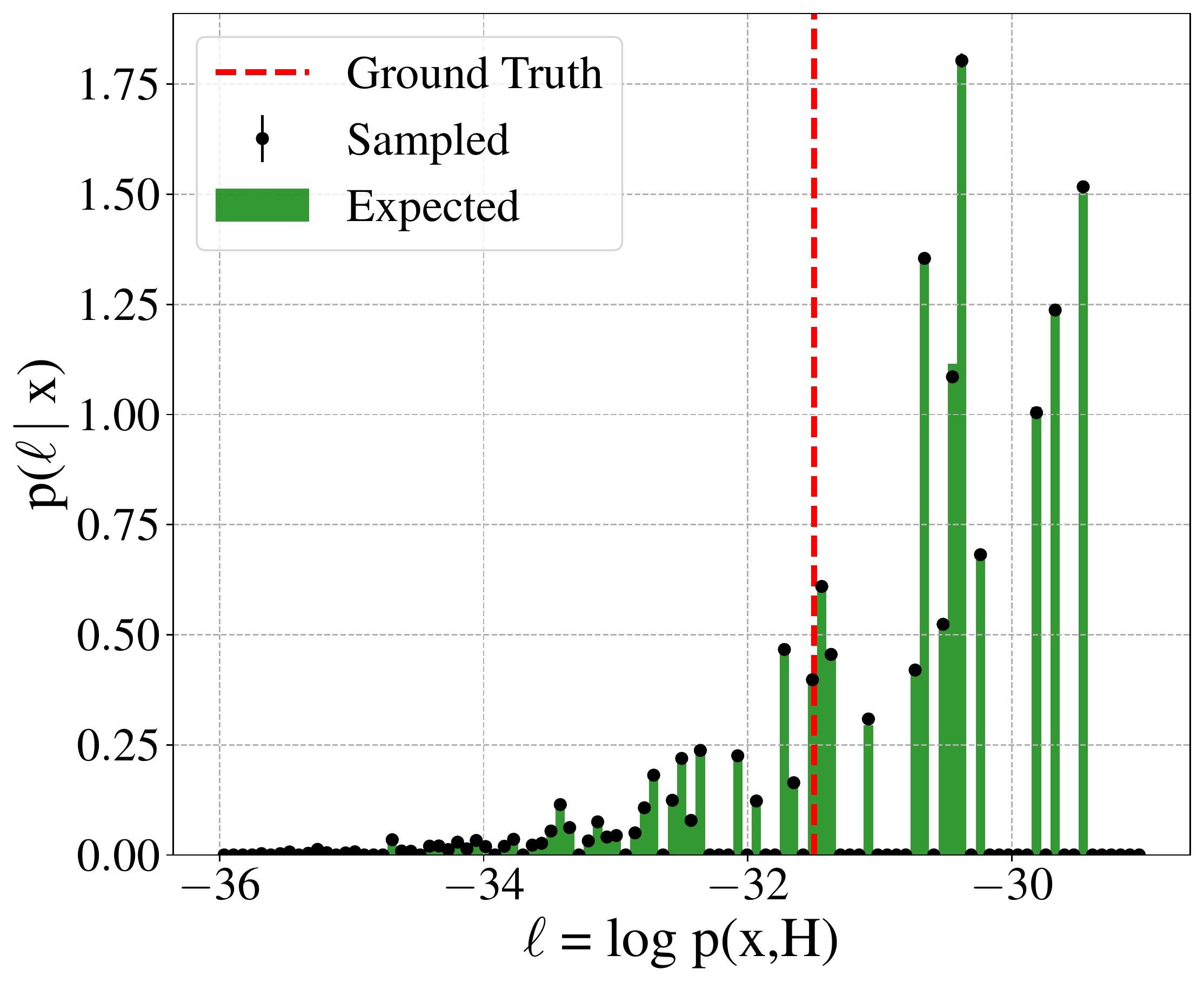}
        \caption{\small{Comparison of the posterior distribution for  a specific jet with five leaves for sampling $10^5$ hierarchies using Alg \ref{alg:sampling_function_trellis} (black dots with small error bars) and expected posterior distribution (in green). The plots show the discrete nature of the distribution.
        The log likelihood for the ground truth tree is a vertical dashed red line.
        % , which happens to be near the MAP value in this case.
        }}
        \label{fig:posteriorJets}
    \end{minipage}
\end{figure}

Next, we review new implementations of greedy and beam search algorithms to cluster jets based on the joint likelihood of the jet binary tree in Ginkgo. The goal is to obtain the maximum likelihood estimate (MLE) or MAP for the latent structure of a jet.
In this approach, the tree latent structure $\tree$ is fixed by the algorithm. Greedy simply chooses the pairing of nodes that locally maximizes the likelihood at each step, whereas beam search maximizes the likelihood of multiple steps before choosing the latent path. The current implementation only takes into account one more step ahead, with a beam size given by $\frac{N(N-1)}{2}$, with $N$ the number of jet constituents to cluster. Also, when two or more clusterings had an identical likelihood value, only one of them was kept in the beam, to avoid counting multiple times the different orderings of the same clustering (see \cite{boyles2012time} for details about the different orderings of the internal nodes of the tree). This approach significantly improved the performance of beam search in terms of finding the MAP tree. 
\vspace{-0.1cm}
\paragraph{Results}
In this section we show results for a jet physics dataset of 5000 Ginkgo \cite{ToyJetsShowerPackage} jets with a number of leaves between 5 and 10, and we refer to it as Ginkgo510. We start by comparing in Table \ref{tab:meandiff} the mean difference among the MAP values for the hierarchies log likelihood obtained with the full trellis, beam search and greedy algorithms. We see that the likelihood of the trees increases from greedy to beam search to the trellis one,
as expected.  
% We use beam search as a baseline to estimate the MAP value,
% which typically has a good performance for trees with up to about 10 leaves, but as we see in Table \ref{tab:meandiff}, the trellis MAP value is greater. 
Next, in Figure \ref{fig:ZvsMLEscatter} we show the  partition function versus the MAP hierarchy for each set of leaves in Ginkgo510 dataset. It is interesting to note that there seems to be a correlation between $Z$ and the Trellis MAP. 
% We want to emphasize that the implementation of the trellis algorithm allows us to access the partition function.

\begin{SCtable}
\centering
\caption{\small{
Mean and standard deviation for the difference in log likelihood for the MAP tree found by algorithms indicated by the row and column heading on the Ginkgo510 dataset.
%Mean and standard deviation for the difference in the MAP values of the hierarchies likelihood (logarithmic scale) over Ginkgo510 dataset.
}}
\vspace{-0.1cm}
\label{tab:meandiff}
%\vskip -0.5in
\begin{small}
\begin{sc}
\begin{tabular}{lcccr}
\toprule
  & \bf Beam Search & \bf Greedy \\
\midrule
\bf Trellis      & 0.4 $\pm$ 0.5 & 1.5 $\pm$ 1.1 \\
\bf Beam Search   &  & 1.1 $\pm$ 1.1\\
\bottomrule
\end{tabular}
\end{sc}
\end{small}
\vspace{-0.5cm}
\end{SCtable}

% \begin{figure}[ht]
% \vspace{-0.3cm}
%     \centering
%     % \begin{minipage}{0.35\textwidth}
%         \centering
%         \includegraphics[width=0.5\columnwidth]{figs/ZvsMLEscatter.pdf}
%         \vspace{-0.2cm}
%         \caption{\small{Scatter plot of the partition function $Z$  vs. the trellis MAP value $\ell$ for Ginkgo510 dataset, with up to 10 leaves (jet constituents). The color indicates the number of leaves of each hierarchical clustering. There appears to be a correlation between $Z$ and the MAP values.
%         }}
%         \label{fig:ZvsMLEscatter}
%         \vspace{-0.3cm}
%     % \end{minipage}
% \end{figure}

% \begin{figure}
% \vspace{-0.2cm}
%     % \begin{minipage}{0.35\textwidth}
%         \centering
%         \includegraphics[width=0.6\columnwidth]{figs/posteriorSample100000_5_6.pdf}
%         \vspace{-0.2cm}
%         \caption{\small{Comparison of the posterior distribution for  a specific jet with five leaves for sampling $10^5$ hierarchies using Alg \ref{alg:sampling_function_trellis} (black dots with small error bars) and expected posterior distribution (in green). The plots show the discrete nature of the distribution.
%         The log likelihood for the ground truth tree is a vertical dashed red line.
%         }}
%         \label{fig:posteriorJets}
%         \vspace{-0.4cm}
%     % \end{minipage}
% \end{figure}
\vspace{-.1cm}

Next, we show an implementation of the sampling procedure introduced in section \ref{sec:Treesampling}. 
% We compare the sampled posterior distribution with respect to the expected one, conditioned on a set of five leaves.
We compare in Figure \ref{fig:posteriorJets} the results from sampling $10^5$ hierarchies (black dots) and the expected distribution\footnote{The expected posterior is defined as the probability density function of each possible hierarchy. In principle, this could be obtained by taking the ratio of the likelihood of each hierarchy with respect to the partition function $Z$. 
We opt to take an approximate approach, as follows. If we sample enough number of times, we would expect each possible hierarchy to appear at least once. Thus, as a proof of concept, we sample $10^5$ hierarchies for a set of five leaves (88 different hierarchies), keep only one of them for each unique likelihood value and normalize by $Z$ and bin size. We show this result in the histogram labeled as Expected (green) in Figure \ref{fig:posteriorJets}. } (green) for the likelihood of each hierarchy.
There is an excellent agreement between the sampled and the expected distributions. Here we showed, for illustrative purposes, a way to estimate the posterior distribution using our sampling procedure. However, we want to emphasize that the key contribution of our procedure is that it allows to sample hierarchies from the exact true posterior distribution, i.e. sample a hierarchy according to its probability.

\begin{figure}
\vspace{-0.2cm}
\centering
\includegraphics[width=0.5\columnwidth]{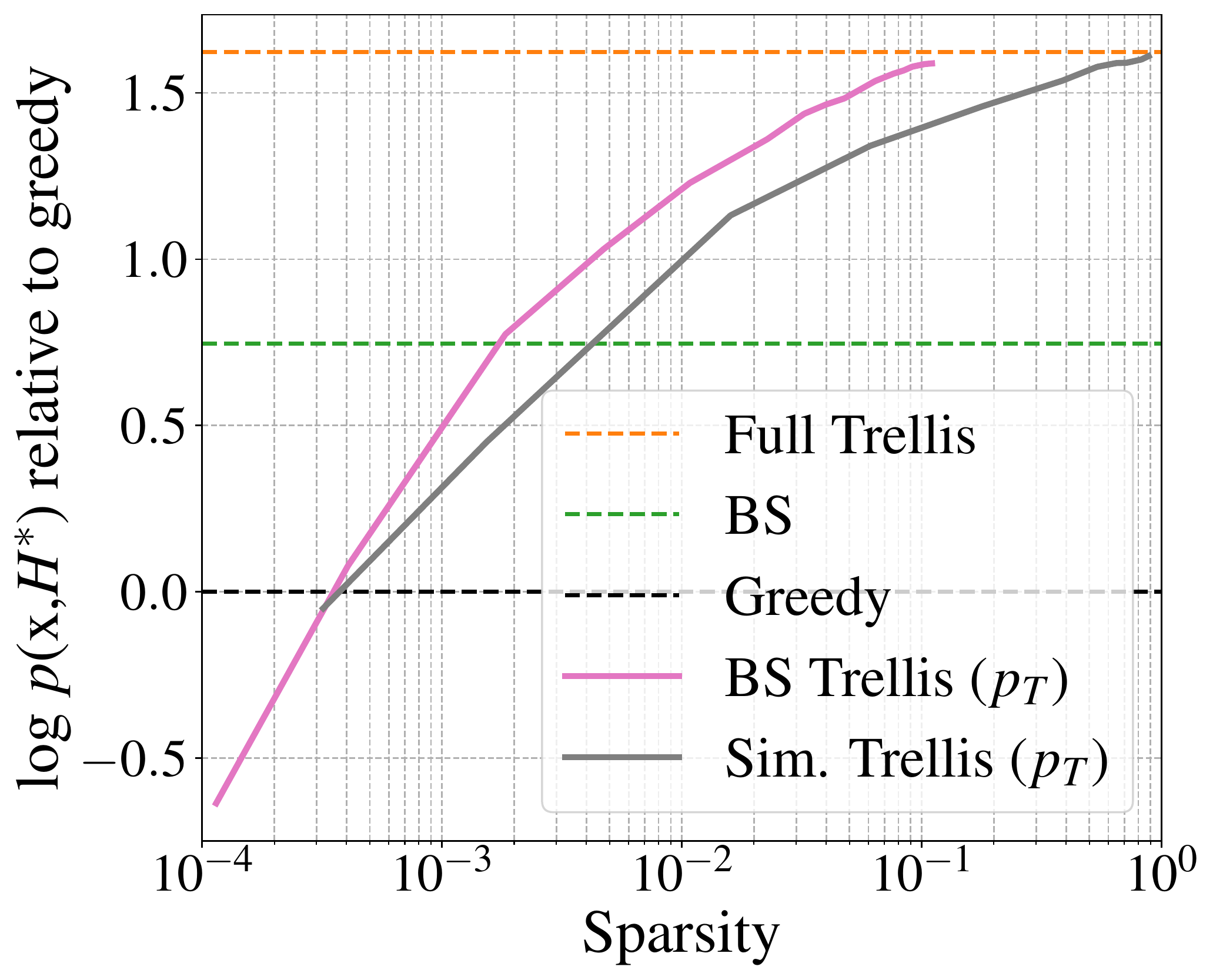}
\vspace{-0.3cm}
\caption{\small{Trellises MAP hierarchy log likelihood (values are relative to the greedy algorithm) vs their sparsity. Each value corresponds to the mean over 100 trees of a test dataset. We show the Simulator (Sim.) and the Beam Search (BS) trellises. In both cases, we present the trellis obtained by ordering the leaves of the input trees in increasing norm of their momentum vector $\vec{p} \in \RR^3$ (see the probabilistic model description of section \ref{exp:jets} for more details). We compare sparse trellises with other orderings of the leaves in Appendix $\S$ \ref{sparsetrellis}.
We add the values of the exact trellis, beam search and greedy algorithms. The BS trellis approaches the performance of the full one for a smaller sparsity index than the Sim. Trellis. 
% Also, the results shown are obtained from running the sparse trellises on a different dataset than the one used to , after having being pre-built on a different one
Also, the sparse trellises are pre-built and then run on new datasets (test), which is why BS performs better than BS trellis sometimes.}
% In each case, we present the trellis obtained by ordering the leaves of the input trees in increasing norm of their momentum vector $\vec{p} \in \RR^3$ (see the probabilistic model description of section \ref{exp:jets} for more details) or randomly, labeled as $p_T$ or {\it rand} respectively.
% We add the values of the exact trellis, beam search and greedy algorithms for comparison. The BS trellis approaches the performance of the full one for a smaller sparsity index than the Sim. Trellis. 
}
\label{fig:MAPvsSparsity}
\vspace{-0.3cm}
\end{figure}

Finally, as a proof of concept, we show in Figure \ref{fig:MAPvsSparsity} the performance of the sparse trellis to calculate the MAP values on a set of 100 Ginkgo jets with 9 leaves. We chose a dataset of 9 elements to be able to easily compare the performance of the sparse and full trellises. However, the sparse trellis can be applied to larger datasets. Even though beam search has a good performance for trees with a small number of leaves, we see that both sparse trellises quickly improve over beam search, with a sparsity index of only about 2\%. 
% For these values of the sparsity index, the trellis is efficient and fast to run. 
Both sparse trellises approach the performance of the exact one, but the BS trellis does it sooner.

% #################################
\vspace{-0.2cm}
\subsection{Cancer Genomics}
\label{sec:genomics}
\vspace{-0.2cm}
\paragraph{Background}
Hierarchical clustering is a common clustering approach for gene expression data~\citep{sorlie2001gene}. 
However, standard hierarchical clustering uses a greedy agglomerative or divisive heuristic to build a tree.
It is not uncommon to have a need for clustering a small number of samples in cancer genomics studies.
An analysis of data available from \href{https://clinicaltrials.gov}{\url{https://clinicaltrials.gov}} shows that the median sample size for 7,412 completed phase I clinical trials involving cancer is only 30.

\vspace{-0.1cm}
\paragraph{Probabilistic Model}
In this case we are given a dataset of vectors indicating the level of gene expressions which are endowed with pairwise
affinities that are both positive and negative. 
% We define the energy of a pair of sibling nodes in the tree to be the sum of the positive edges not crossing the cut\footnote{Crossing the cut means across subcluster edges, i.e. edges from one subcluster to another.}, minus the sum of the negative edges crossing the cut,
We define the energy of a pair of sibling nodes in the tree to be the sum of the positive edges from elements in one child to elements in the other one, minus the negative edges between two elements in the same child.

\vspace{-0.4cm}
{\small
\begin{align}
    \EfunS(X_i, X_j) = & \exp(-\beta E(X_i,X_j)) \\
    E(X_i,X_j) =& \sum_{\mathclap{x_i,x_j \in X_i \times X_j}} w_{ij} \II[w_{ij} > 0] - \sum_{\mathclap{\substack{x_i, x_j \in X_i \times X_i, \\ x_i \not = x_j }}} w_{ij} \II[w_{ij} < 0]  - \sum_{\mathclap{\substack{x_i,x_j \in X_j \times X_j, \\ x_i \not = x_j}}} w_{ij} \II[w_{ij} < 0] 
\end{align}
}
\vspace{-0.1cm}
where $w_{ij}$ is the affinity between $x_i$ and $x_j$.
%\craig{don't need -E() since $\EfunS$ includes the negation}
The correlation clustering input can be represented as a complete weighted graph, $G = (V,E)$, where each edge has weight $w_{uv} \in [-1,1], \forall (u,v) \in E$. 
The goal is to construct a clustering of the nodes that maximizes the sum of positive within-cluster edge weights minus the sum of all negative across-cluster edge weights. 
This energy is the correlation clustering objective \cite{bansal2004correlation}.
\vspace{-0.1cm}
\paragraph{Data and Methods}
Here, we compare a greedy agglomerative clustering to our exact MAP clustering tree using the Prediction Analysis of Microarray 50 (\texttt{pam50}) gene expression data set.
The \texttt{pam50} data set ($n=232$, $d=50$) is available from the UNC MicroArray Database \citep{UNCMicro}.
It has intrinsic subtype annotations for 139 of the 232 samples.
Missing data values (2.65\%) were filled in with zeros.
We drew a stratified sample of the total data set with two samples from each known intrinsic subtype and two samples from the unknown group.  
% See Appendix $\S$ \ref{app:genomics} for more details.
\begin{figure}
\vspace{-0.2cm}
  \centering
    \includegraphics[width = 0.7\textwidth]{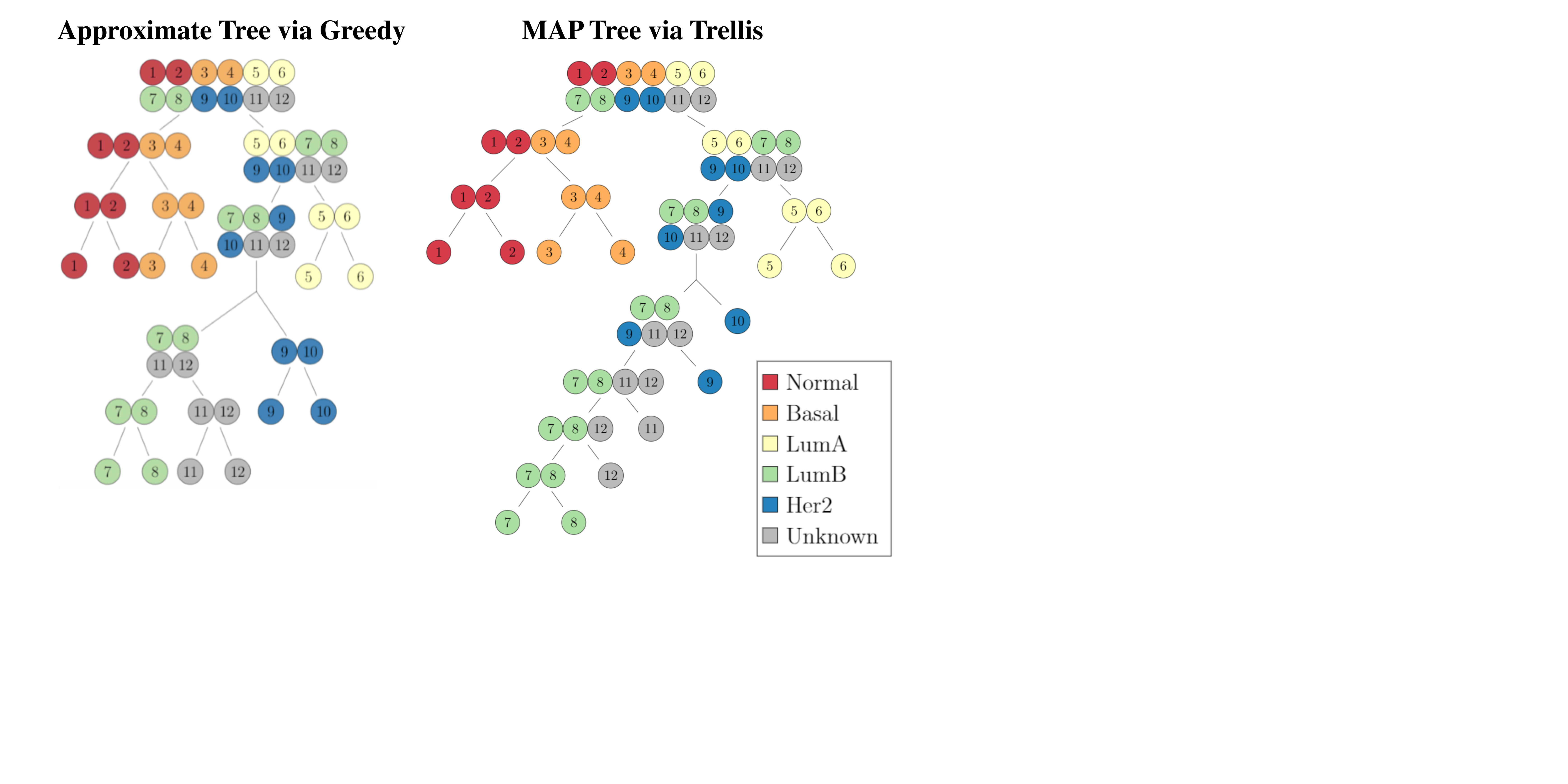}
    \vspace{-0.3cm}
    {\caption{\small{\textbf{Cancer Genomics}. Comparison of trees from greedy hierarchical clustering (left) and exact MAP clustering using the trellis (right) on the subsampled \texttt{pam50} data set. The colors indicate subtypes of breast cancer (grey if unknown). Though both appear to assign unknown samples to LumB, the right tree positions the unknown samples closer to the Her2 samples.}}\label{fig:pamAssignment}}
    \vspace{-0.5cm}
\end{figure}

\vspace{-0.1cm}
\paragraph{Results}
Figure \ref{fig:pamAssignment} displays the greedy hierarchical clustering tree and the MAP tree with transformed weights 
% (see Appendix $\S$ \ref{app:genomics} for more details)
for the twelve samples selected from the pam50 dataset.
% \footnote{To allow more balanced weights among vertices, we transform the weights using $w_{uv}' = \tan \left(w_{uv} \frac{\pi}{2} \right)$ and then subtracted the average of all $w_{uv}'$.} 
(The correlations among subsampled \texttt{pam50} ($n=12$) data set are all positive.) 
The main difference between these trees is in the split of the subtree including LumB, HER2, and unknown samples.
The greedy method splits HER2 from LumB and unknown, while the MAP tree shows a different topology for this subtree.
For the MAP solution, we note that the subtree rooted at $\{7,8,9,10,11,12\}$ is consistent.
All of the correlation coefficients among this cluster are positive, so the optimal action is to split off the item with the smallest (positive) correlation coefficient.

\vspace{-0.2cm}
\section{Related Work}
\label{related work}
\vspace{-0.2cm}
Modeling distributions over tree structures has been the subject of 
a large body of work. 
% Despite this, modeling distributions over tree structures has been the subject of a large body of work, including various types of Bayesian non-parametric models, e.g., \cite{knowles2011pitman, hu2013binary, blei2010nested, paisley2014nested, ghahramani2010tree}.  These methods only support using parametric distributions to define emission probabilities and do not support the general family of potential functions as in our method. Bootstrapping methods, such as \cite{suzuki2006pvclust}, represent uncertainty in hierarchical clustering, however they approximate statistics of interest through repeatedly (re-)sampling from the empirical distribution. Recent work \cite{kappes2015probabilistic, kohonen2016computing, greenberg2018compact} has studied distributions over flat clusterings and showed that dynamic programming can be used to both efficiently compute the partition function as well as find the MAP clustering \cite{van2004improving, NIPS2018_8081}. However, analogous results for hierarchical clustering were not known.
Bayesian non-parametric models typically define a posterior distribution over tree structures given data such as
diffusion trees coalescents, and others \cite[inter alia]{neal2003density,teh2008bayesian}.
% \cite{neal2003density,knowles2011pitman},
% coalescent models \cite{teh2008bayesian,boyles2012time,hu2013binary}, and in the case of grouped data, the nested Chinese restaurant processes \cite{blei2010nested} and nested hierarchical Dirichlet processes \cite{paisley2014nested}. 
% Other models, such as tree structured nested sticking breaking, provide a distribution over a different class of tree structures, one for which data can sit at internal nodes \cite{ghahramani2010tree}. 
These methods, while providing a distribution over trees,  only support using parametric distributions to define emission probabilities and do not support the general family of potential functions 
used in our approach, which can use any scoring function to define the distribution. Factor graph-based distributions over tree structures such as \cite{wick2012discriminative} on the other hand support a flexible class of distributions over tree structures as in our approach. However inference in factor graph models as well as many of the Bayesian non-parameteric models is typically approximate or performed by sampling methods. This lends in practice to approximate MAP solutions and distributions over tree structures. Exact methods like the one proposed in this paper have not, to our knowledge, been proposed. 

\cite{dasgupta2016cost} defines a cost function for hierarchical clustering. Much work has been done to develop approximate solution methods and related objectives \cite[inter alia]{moseley2017approximation}. 
% \cite{roy2017hierarchical,cohen2019hierarchical,charikar2017approximate,moseley2017approximation,cohen2017hierarchical,charikar2019hierarchical}.

Bootstrapping methods, such as \cite{suzuki2006pvclust}, represent uncertainty in hierarchical clustering.  Unlike our approach, bootstrapping methods approximate statistics of interest through repeatedly (re-)sampling from the empirical distribution.

Work on exact inference and exact distributions over flat clusterings \cite{NIPS2018_8081}, provides the foundation of our dynamic programming approach. Other work on exact flat clustering uses fast convolutions via the Mobius transform and Mobius inversion \cite{kohonen2016computing}. \cite{kappes2015probabilistic} produce approximate distributions over flat clusterings using Perturb and MAP \cite{papandreou2011perturb}.

% \sebastian{Recent work studies gradient-based hierarchical clustering \cite{10.1145/3292500.3330997,chami2020trees}. In this work, hierarchies are parametrized by embeddings of their elements in a hyperbolic space, and the least common ancestor is analog to a geodesic in that space. Gradient based optimization is implemented on a differentiable variant of Dasgupta's optimization cost \cite{dasgupta2016cost} in \cite{chami2020trees}.}

Orthogonal to our work on uncertainty in hierarchical clustering, recent work has proposed continuous representations of trees for hierarchical clustering \cite{10.1145/3292500.3330997,chami2020trees}. This work represents uncertainty of child-parent assignments by considering the distance between two nodes in embedding space. We note that the distribution over trees used in these papers does not directly correspond to the energy-based distribution proposed in our work. 
\vspace{-0.2cm}
\section{Conclusion}
\vspace{-0.2cm}
This paper describes a trellis data structure and dynamic-programming algorithm to efficiently compute and sample from probability distributions over hierarchical clusterings. 
Our method improves upon the computation cost of brute-force methods from $\exactNumClusterings$ to sub-quadratic
in the substantially smaller powerset of $N$, which is super-exponentially more efficient. We demonstrate our methods' utility on jet physics and cancer genomics datasets, as well as a dataset related to Dasgupta's cost \cite{dasgupta2016cost}, and show its improvement over approximate methods.
Also, for larger datasets where the full trellis implementation becomes infeasible, we introduce a sparse trellis that compares well to other benchmarks. Finally, our methods allow to sample hierarchies from the exact true posterior distribution without enumerating all possible ones, i.e. sample a hierarchy according to its probability.
% \sebastian{How about deleting this:
% Finally, though there are multiple approaches to sample from the posterior distribution, such as MCMC techniques, they could be very expensive while our sampling method has a bounded computation time.}
\vspace{-0.2cm}

\section*{Acknowledgements}

Kyle Cranmer and Sebastian Macaluso are supported by the National Science Foundation under the awards ACI-1450310 and OAC-1836650 and by the Moore-Sloan data science environment at NYU. Patrick Flaherty is supported in part by NSF HDR TRIPODS award 1934846. Andrew McCallum and Nicholas Monath are supported in part by the Center for Data Science and the Center for Intelligent Information Retrieval, and in part by the National Science Foundation under Grant No. NSF-1763618. Any opinions, findings and conclusions or recommendations expressed in this material are those of the authors and do not necessarily reflect those of the sponsor.

\bibliographystyle{unsrt}
\bibliography{references}
% \clearpage
% \input{AA_appendix}
\newpage
\appendix
\clearpage
\onecolumn
\section{Appendix}
%\subsection{Proof of Fact %\ref{thm:recursive-partition}}
%\input{tex/partition_function_proof.tex}

\subsection{Proof of Proposition \ref{thm:recursive-partition}}
\label{thm:recursive-partition-proof}
\begin{proof}
Given a dataset $\dataset$, pick an element $x \in \dataset$. We consider all possible $\Omega$ clusters $\dataset^\omega_L$ in $\children(\dataset)_{x}$. Given $\dataset^\omega_L$, then $\dataset_R^\omega$ is fixed so as to satisfy $\dataset_L^\omega \bigcup \dataset_R^\omega = \dataset$ and  $\dataset_L^\omega \bigcap \dataset_R^\omega = \emptyset$. 
We want to show that the partition function $Z(X)$ can be written recursively in terms of $Z(\dataset_L^\omega)$ and $Z(\dataset_R^\omega)$.

The partition function is defined as the sum of the energies of all possible hierarchical clusterings $\treeset_\dataset = \{\tree^m\}_{m=1}^M$,
{\small
\begin{align}\label{eq:Zdef}
Z(\dataset) & = \sum_{m=1}^M \,\Efun(\tree^m(\dataset))
             =  \sum_{m=1}^M \, \EfunS(\dataset_L^m,\dataset_R^m)\,\, \Efun(\tree^m(\dataset_L^m)) \, \, \Efun(\tree^m(\dataset_R^m))
\end{align}
}%
where $\dataset_L^m \bigcup \dataset_R^m = \dataset$,  $\dataset_L^m \bigcap \dataset_R^m = \emptyset$. Also, $\tree^m(\dataset_{L}^m)$ and $\tree^m(\dataset_{R}^m)$ are the sub-hierarchies in $\tree^m$ that are rooted at $\dataset_{L}^m$ and $\dataset_{R}^m$, respectively. %\craig{I don't see any $X$ with a double subscript, i.e., $\dataset_{L,R}^m$, in the above equation.}
Next, we rewrite Eq. \ref{eq:Zdef} grouping together all the hierarchies $\tree^i$ that have the same clusters $\{\dataset_L^m,\dataset_R^m\}$ \footnote{The cluster trellis provides an exact solution conditioned on the fact that the domain of the linkage function is the set of pairs of clusters, and not pairs of trees.},
{\small
\begin{align}\label{eq:Zdef2}
Z(\dataset) & = \sum_{\omega=1}^\Omega \, \EfunS(\dataset_L^\omega,\dataset_R^\omega)\,\, \sum_{j=1}^J \,\Efun(\tree^j(\dataset_L^\omega)) \,\, \sum_{k=1}^K \,\Efun(\tree^k(\dataset_R^\omega)) 
 = \sum_{\omega=1}^\Omega \, \EfunS(\dataset_L^\omega,\dataset_R^\omega)\,\, Z(\dataset_L^\omega)\,\, Z(\dataset_R^\omega)
\end{align}
}%
with $M = \Omega \cdot J \cdot K$,  $J = (2 |\dataset_{L}^{\omega}| - 3)!!$, and $K = (2 |\dataset_{R}^{\omega}| - 3)!!$ for a full trellis. Thus, $Z(X)$ of a cluster $X$ can be written recursively in terms of the partition function of the sub-clusters of X \footnote{Note that for each singleton ${x_i}$, we have $Z(x_i)=1$.}.

% Given a dataset $\dataset$, pick $x \in \dataset$. We consider all possible $\Omega$ clusters $\dataset^\omega_L \subset \dataset$ that contain $x$. Given $\dataset^\omega_L$, then $\dataset_R^\omega$ is fixed so as to satisfy $\dataset_L^\omega \bigcup \dataset_R^\omega = \dataset$ and  $\dataset_L^\omega \bigcap \dataset_R^\omega = \emptyset$.

\end{proof}

\subsection{Proof of Theorem \ref{thm:sampling}}
\label{proof:sampling-method}
\begin{proof}
We want to show that drawing samples of trees using 
Algorithm \ref{alg:sampling_function_trellis} gives samples from $P(\tree|\dataset)$. To do this, we show 
that the probability of a tree can be re-written
as the product of probabilities of sampling each split
in the structure. This then directly corresponds to 
the top-down sampling procedure in Algorithm \ref{alg:sampling_function_trellis}.

Recall from Definition \ref{defn:energy_based_hclustering} we have:
{\small \begin{align}
    P(\tree|\dataset) &= \frac{1}{Z({\dataset})}  \prod_{X_L,X_R \in \textsf{sibs}(\tree)} \EfunS(X_L,X_R) 
 \end{align}
 }%
 We can equivalently write this as:
{\small
 \begin{align}\label{eq:s12}
    P(\tree|\dataset) =&  \prod_{X_L,X_R \in \textsf{sibs}(\tree)} \frac{1}{Z({X_L \cup X_R})}\cdot    \EfunS(X_L,X_R) \cdot Z(\dataset_L)\cdot Z(\dataset_R) 
 \end{align}
 }%
 To understand why this can be written this way, observe that for internal nodes the $Z(X_L)$ and $Z(X_R)$ terms will be cancelled out by corresponding terms in the product for the children of $X_L$ or $X_R$. To see this we can write out the product for three pairs of nodes $X_L$, $X_R$ and their children $X_{LL}, X_{LR}$ and $X_{RL}$ and $X_{RR}$ respectively: 
 
{\footnotesize
\begin{align}\label{Eq:p2levels}
% p = & \, 
 & \frac{1}{Z(\dataset_p)}\EfunS(\dataset_L,\dataset_R) \,\, Z(\dataset_L) \,\, Z(\dataset_R) \nonumber\\ & \cdot \frac{1}{Z(\dataset_L)}\EfunS(\dataset_{LL},\dataset_{LR}) \,\, Z(\dataset_{LL}) \,\, Z(\dataset_{LR}) \nonumber\\  & \cdot \frac{1}{Z(\dataset_R)}\EfunS(\dataset_{RL},\dataset_{RR}) \,\, Z(\dataset_{RL}) \,\, Z(\dataset_{RR})
\end{align}
}%

Recall that for the pair of siblings that are the children of the root, the $\frac{1}{Z(X_L \cup X_R)}$ term will not be cancelled out and corresponds exactly to $\frac{1}{Z(X)}$. 

Next, we observe that Eq. \ref{eq:s12} can be re-written in terms of Equation \ref{eq:levelSampling} which defines $p(X_L | X_L \cup X_R)$:
 \begin{align}\label{eq:correspond_sample}
    P(\tree|\dataset) &=  \prod_{X_L,X_R \in \textsf{sibs}(\tree)} p(X_L | X_L \cup X_R) 
 \end{align}

Algorithm \ref{alg:sampling_function_trellis} applies Eq. \ref{eq:levelSampling} recursively in a top-down manner using a series of splits which have a probability that directly corresponds to the product  of terms in Eq. \ref{eq:correspond_sample}.

\end{proof}

\subsection{Proof of Lower Bound on Number of Trees}
\label{proof:tree_bound}
% \craig{Do we still reference and/or need this?}
%$\NumHierachicalClusters$ trees
The number of trees on $N$ leaves is given exactly by $\frac{\prod_{i=1}^{N-1}{m_i}}{(N-1)!}\prod_{i=2}^{N}{\binom{i}{2}}$, where $m_i$ is the number of internal nodes in the subtree rooted at node $i$ \cite{boyles2012time}. Since $\prod_{i=2}^{N}{\binom{i}{2}} = \frac{N!^2}{N * 2^{N-1}}$, this makes the number of trees on $N$ leaves $\frac{\prod_{i=1}^{N-1}{m_i}}{(N-1)!}\frac{N!^2}{N * 2^{N-1}} = \frac{\prod_{i=1}^{N-1}{m_i} * N*N!}{N * 2^{N-1}} = \frac{\prod_{i=1}^{N-1}{m_i} * N!}{2^{N-1}}$. %Product[Binomial[i, 2], {i, 2, N}] = (2^(1 - N) N(N!)^2)/N 
The smallest conceivable value for $\prod_{i=1}^{N-1}{m_i} = \omega(N)$, which gives us the bound on the number of trees to be
$\NumHierachicalClusters$, as desired.

Note that this is a loose lower-bound, and that it could be improved upon as follows: say a hierarchical clustering is a \emph{caterpillar clustering} is every internal node in the underlying tree has two children and the set associated with one of those children as size one. There are $n!/2$ caterpillar clustering. To see this, note that the $i$th level (where the root is level 1) of a caterpillar clustering has exactly one leaf for $i=2, \ldots, n-1$. There are $n(n-1)\ldots 3=n!/2$ choices for the corresponding singleton sets.

Note, however, that there is a closed form expression for the exact number of unordered hierarchies given by $ a(N) = \exactNumClusterings$, with n the number of singletons (see \cite{callan2009combinatorial,DaleMoonCatalanSets} for more details and proof).

%\sebastian{We want to mention that there is a closed form expression for the exact number of unordered hierarchies given by $ a(N) = \exactNumClusterings$, with n the number of singletons. }

\subsection{Correctness Proof of Marginal Algorithms} 
\label{proof:marginal-formulas}
\subsubsection{Sub-Hierarchy Marginal}
For a given sub-hierarchy rooted at $\dataset_i$, i.e., $\tree_i \in \treeset(\dataset_i)$, the marginal probability is defined as ${P(\tree_i|\dataset) = {\sum_{\tree \in A(\tree_i)}} P(\tree|\dataset) }$
, where $A(\tree_i) = \{\tree : \tree \in \treeset(\dataset) \land \tree_i \subset \tree\}$, and $\tree_i \subset \tree$ indicates that $\tree_i$ is a subtree of $\tree$. We can rewrite ${\sum_{\tree \in A(\tree_i)}} P(\tree|\dataset)$ as ${\sum_{\tree \in A(\tree_i)}} \EfunT(\tree(\dataset)) / Z $, which gives us: 
{\small
\begin{align}%\label{eq:marge-def}
    P(\tree_i|\dataset) = {\sum_{\tree \in A(\tree_i)}} P(\tree|\dataset) = {\sum_{\tree \in A(\tree_i)}} \frac{\EfunT(\tree(\dataset))}{Z} = \frac{Z_{\tree_i}(\dataset)}{Z} 
\end{align}
}%

where $Z_{\tree_i}(\dataset) = {\sum_{\tree \in A(\tree_i)}} \EfunT(\tree(\dataset))$, the sum of potential values for all the hierarchies containing the sub-hierarchy $\tree_i$. This gives us 
{\small \begin{align}\label{eq:Z-H_idef}
Z_{\tree_i}(\dataset) & = \sum_{m=1}^{|A(\tree_i)|} \,\Efun(\tree^m(\dataset))\nonumber \\
            & =  \sum_{m=1}^{|A(\tree_i)|} \, \EfunS(\dataset_L^m,\dataset_R^m)\,\, \Efun(\tree^m(\dataset_L^m)) \, \, \Efun(\tree^m(\dataset_R^m))
\end{align}
}%

where $\dataset_L^m \bigcup \dataset_R^m = \dataset$,  $\dataset_L^m \bigcap \dataset_R^m = \emptyset$. Also, $\tree^m(\dataset_{L}^m)$ and $\tree^m(\dataset_{R}^m)$ are the sub-hierarchies in $\tree^m$ that are rooted at $\dataset_{L}^m$ and $\dataset_{R}^m$, respectively. 
Next, we rewrite Eq. \ref{eq:Z-H_idef} grouping together all the hierarchies $\tree^i$ that have the same clusters $\{\dataset_L^m,\dataset_R^m\}$.
Note that $\tree_i \subset \tree$, implies $\dataset_i \subseteq \dataset_L$ or  $\dataset_i \subseteq \dataset_R$.  Assume W.L.O.G. that $\dataset_i \subseteq \dataset_L$.
{\small
\begin{align}\label{eq:marge-def}
   Z_{\tree_i}(\dataset) & = \sum_{\omega=1}^\Omega \, \EfunS(\dataset_L^\omega,\dataset_R^\omega)\,\, \sum_{j=1}^J \,\Efun(\tree^j(\dataset_L^\omega)) \,\, \sum_{k=1}^K \,\Efun(\tree^k(\dataset_R^\omega)) \nonumber \\
& = \sum_{\omega=1}^\Omega \, \EfunS(\dataset_L^\omega,\dataset_R^\omega)\,\, Z_{\tree_i}(\dataset_L^\omega)\,\, Z(\dataset_R^\omega)
    \nonumber \\
\end{align}
}%
with $|A(\tree_i)|= \Omega \cdot J \cdot K$, $J = |\{\treeset(X_L):  \dataset_i \subseteq \dataset_L\}|$, $K = |\{\treeset(\dataset_R)\}|$ and setting $Z_{\tree_i}(\dataset_i) = \Efun(\tree(\dataset_i))$. 

\subsubsection{Subset Marginal}

For a given cluster $\dataset_i$, the marginal probability is defined as ${P(\dataset_i|\dataset) = {\sum_{\tree \in A(\dataset_i)}} P(\tree|\dataset) }$
, where $A(\dataset_i) = \{\tree : \tree \in \treeset(\dataset) \land \dataset_i \subset \tree\}$, and $\dataset_i \subset \tree$ indicates that cluster $\dataset_i$ is contained in $\tree$. We can rewrite ${\sum_{\tree \in A(\dataset_i)}} P(\tree|\dataset)$ as ${\sum_{\tree \in A(\dataset_i)}} \EfunT(\tree(\dataset)) / Z $, which gives us: 
{\small 
\begin{align}\label{eq:marge-def2}
    P(\dataset_i|\dataset) = {\sum_{\tree \in A(\dataset_i)}} P(\tree|\dataset) = {\sum_{\tree \in A(\dataset_i)}} \frac{\EfunT(\tree(\dataset))} {Z} = \frac{Z_{\dataset_i}(\dataset)}{Z} 
\end{align}
}%

where $Z_{\dataset_i}(\dataset) = {\sum_{\tree \in A(\dataset_i)}} \EfunT(\tree(\dataset))$, the sum of potential values for all the hierarchies containing the cluster $\dataset_i$. This gives us 
{\small
\begin{align}\label{eq:Z-H_idef2}
Z_{\dataset_i}(\dataset) & = \sum_{m=1}^{|A(\dataset_i)|} \,\Efun(\tree^m(\dataset))
            =  \sum_{m=1}^{|A(\dataset_i)|} \, \EfunS(\dataset_L^m,\dataset_R^m)\,\, \Efun(\tree^m(\dataset_L^m)) \, \, \Efun(\tree^m(\dataset_R^m))
\end{align}
}%
where $\dataset_L^m \bigcup \dataset_R^m = \dataset$,  $\dataset_L^m \bigcap \dataset_R^m = \emptyset$. Also, $\tree^m(\dataset_{L}^m)$ and $\tree^m(\dataset_{R}^m)$ are the sub-hierarchies in $\tree^m$ that are rooted at $\dataset_{L}^m$ and $\dataset_{R}^m$, respectively. 
Next, we rewrite Eq. \ref{eq:Z-H_idef2} grouping together all the hierarchies $\tree^i$ that have the same clusters $\{\dataset_L^m,\dataset_R^m\}$.
Note that $\dataset_i \subset \tree$, implies $\dataset_i \subseteq \dataset_L$ or  $\dataset_i \subseteq \dataset_R$.  Assume W.L.O.G. that $\dataset_i \subseteq \dataset_L$.
{\small 
\begin{align}\label{eq:marge-def}
   Z_{\dataset_i}(\dataset) & = \sum_{\omega=1}^\Omega \, \EfunS(\dataset_L^\omega,\dataset_R^\omega)\,\, \sum_{j=1}^J \,\Efun(\tree^j(\dataset_L^\omega)) \,\, \sum_{k=1}^K \,\Efun(\tree^k(\dataset_R^\omega)) \nonumber \\
& = \sum_{\omega=1}^\Omega \, \EfunS(\dataset_L^\omega,\dataset_R^\omega)\,\, Z_{\dataset_i}(\dataset_L^\omega)\,\, Z(\dataset_R^\omega)
    \nonumber \\
\end{align}
}%
with $|A(\tree_i)|= \Omega \cdot J \cdot K$, $J = |\{\treeset(X_L):  \dataset_i \subseteq \dataset_L\}|$, $K = |\{\treeset(\dataset_R)\}|$, and setting $Z_{\dataset_i}(\dataset_i) = Z(\dataset_i)$. 

\subsection{Proof of MAP Time Complexity}
\label{proof:map-complexity}

The MAP tree is computed for each node in the trellis, and due to the order of computation, at the time of computation for node $i$, the MAP trees for all nodes in the subtrellis rooted at node i have already been computed. Therefore, the MAP tree for a node with $i$ elements can be computed in $2^i$ steps (given the pre-computed partition functions for each of the node's descendants), since the number of nodes for the trellis rooted at node i (with i elements) corresponds to the powerset of i. There are $\binom{n}{i}$ nodes of size $i$, making the total computation $\sum_{i=1}^{N}2^{i}\binom{N}{i} = 3^N-1$.

\begin{algorithm}[tb]
\caption{\texttt{MAP}$(\dataset)$}
\label{alg:map_trellis}
\begin{algorithmic}
    \IF{$\EfunT(\dataset)$ set}
    \STATE \textbf{return} $\EfunT(\dataset),\Xi(\dataset)$
    \ENDIF
    \STATE Pick $x_i \in \dataset$ 
    \STATE $\EfunT(\dataset) \gets -\infty$ 
    \STATE $\Xi(\dataset) \gets \texttt{null}$ \COMMENT{Backpointer to give MAP tree structure.}
    \FOR {$\dataset_i$ in $\children(\dataset)_{x_i}$}
    % \FOR{$\dataset_j$ in $2^{\dataset \setminus \{x_i\}}$} 
    % \STATE $\dataset_i \gets \dataset_j \bigcup \ \{x_i\}$
    \STATE $t \gets \EfunS(\dataset_i, \dataset \setminus \dataset_i) \cdot \phi(\trellisVertex(\dataset_i)) \cdot  \phi(\trellisVertex(\dataset \setminus \dataset_i)) $
    \IF{$\EfunT(\dataset) < t$}
    \STATE $\EfunT(\dataset) \gets t$
    \STATE $\Xi(\dataset) \gets \{\dataset_i, \dataset \setminus \dataset_i\} \cup \Xi(\dataset_i) \cup \Xi(\dataset \setminus \dataset_i)$
    \ENDIF
    \ENDFOR
    \STATE \textbf{return} $\EfunT(\dataset),\Xi(\dataset)$
\end{algorithmic}
\end{algorithm}

\subsection{Proof of Proposition \ref{thm:recursive-map}}
\label{proof:recursive-map}
\begin{proof}
We proceed in a similar way as detailed in Appendix $\S$ \ref{thm:recursive-partition-proof}
, as follows. Given a dataset $\dataset$, pick an element $x \in \dataset$. We consider all possible $\Omega$ clusters $\dataset^\omega_L$ in $\children(\dataset)_{x}$. Given $\dataset^\omega_L$, then $\dataset_R^\omega$ is fixed so as to satisfy $\dataset_L^\omega \bigcup \dataset_R^\omega = \dataset$ and  $\dataset_L^\omega \bigcap \dataset_R^\omega = \emptyset$.
We want to show that the MAP clustering $\Efun(\tree^*(\dataset))$ can be computed recursively  in terms of $\Efun(\tree^*(\dataset_L^\omega))$ and $\Efun(\tree^*(\dataset_R^\omega))$.

The MAP value is defined as the energy of the clustering with maximal energy $\Efun$ among all possible hierarchical clusterings $\treeset_\dataset = \{\tree^m\}_{m=1}^M$,
\begin{align}\label{eq:E_MAPdef}
\Efun(\tree^*(\dataset)) & = \max_{m \in M} \,\Efun(\tree^m(\dataset))\nonumber \\
            & = \max_{m \in M}  \, \EfunS(\dataset_L^m,\dataset_R^m)\,\, \Efun(\tree^m(\dataset_L^m)) \, \, \Efun(\tree^m(\dataset_R^m))
\end{align}
where $\dataset_L^m \bigcup \dataset_R^m = \dataset$,  $\dataset_L^m \bigcap \dataset_R^m = \emptyset$. Also, $\tree^m(\dataset_{L}^m)$ and $\tree^m(\dataset_{R}^m)$ are the sub-hierarchies in $\tree^m$ that are rooted at $\dataset_{L}^m$ and $\dataset_{R}^m$, respectively. As mentioned earlier, the cluster trellis provides an exact MAP solution conditioned on the fact that the domain of the linkage function is the set of pairs of clusters, and not pairs of trees. 
Thus, we can rewrite Eq. \ref{eq:E_MAPdef} grouping together all the hierarchies $\tree^i$ that have the same clusters $\{\dataset_L^m,\dataset_R^m\}$, as follows
\begin{align}\label{eq:Zdef2}
\Efun(\tree^*(\dataset))   = & \max_{\omega \in \Omega} \,\bigg( \EfunS(\dataset_L^\omega,\dataset_R^\omega)\,\,
\max_{j \in J} \,\Efun(\tree^j(\dataset_L^\omega)) \,\, \max_{k \in K} \,\Efun(\tree^k(\dataset_R^\omega)) \bigg) \nonumber \\
 = & \max_{\omega \in \Omega}  \, \EfunS(\dataset_L^\omega,\dataset_R^\omega)\,\, \Efun(\tree^*(\dataset_L^\omega))\,\, \Efun(\tree^*(\dataset_R^\omega))
\end{align}
with $M = \Omega \cdot J \cdot K$. Thus, $\Efun(\tree^*(\dataset))$ of a cluster $X$ can be written recursively in terms of the MAP values of the sub-clusters of X \footnote{Note that for each singleton ${x_i}$, we have $\Efun(\tree^*(x_i))=1$.}.

% Given a dataset $\dataset$, pick $x \in \dataset$. We consider all possible $\Omega$ clusters $\dataset^\omega_L \subset \dataset$ that contain $x$. Given $\dataset^\omega_L$, then $\dataset_R^\omega$ is fixed so as to satisfy $\dataset_L^\omega \bigcup \dataset_R^\omega = \dataset$ and  $\dataset_L^\omega \bigcap \dataset_R^\omega = \emptyset$.

\end{proof}

% \subsection{Proof of Fact \ref{thm:recursive-map}}
% Consider \treestar(\dataset), the hierarchical clustering with the maximal energy over \dataset. Since \treestar(\dataset) is a valid hierarchical clustering, $x$ must be contained in one of the children of root \sebastian{Isn't $x$ contained in one of the children nodes anyways, independently of the MAP tree?}, $X_j$, and the the other child must equal $\dataset \setminus \dataset_j$. Assume toward contradiction that there exists a $\tree(\dataset)$ s.t., 
% $\Efun(\tree(\dataset)) > \Efun(\treestar(\dataset)$. This implies there exists an $X_k$ s.t.,  $\Efun(\dataset_k, \dataset \setminus \dataset_k) * \Efun(\treestar(X_k)) * \Efun(\treestar(\datasetcomplement{X_k})) > \Efun(\dataset_j, \dataset \setminus \dataset_j) * \Efun(\treestar(X_j)) * \Efun(\treestar(\datasetcomplement{X_j}))$, a contradiction.

\subsection{Proofs of Theorem \ref{thm:partition-function-time-complexity} and Corollary \ref{coroll:improvement}}
\label{proof:Z-time-complexity}
The partition function is computed for each node in the trellis, and due to the order of computation, at the time of computation for node $i$, the partition functions for all nodes in the subtrellis rooted at node i have already been computed. Therefore, the partition function for a node with $i$ elements can be computed in $2^i$ steps (given the pre-computed partition functions for each of the node's descendants), since the number of nodes for the trellis rooted at node i (with i elements) corresponds to the powerset of i. %\sebastian{Sorry I don't understand what this exactly means. It seems to me that this way we could just get Z taking the root node, which has N elements?\craig{This is because the $2^i$ value depends on every descendent node also being computed (each of which themselves take $2^k$, where $k$ is the number of elements in the descendent node).  Did adding the above parenthetical help, or is still more explanation needed?}} \sebastian{Could we also add that the $2^i$ is because the trellis data structure runs in time complexity proportional to the powerset of i ?} 
There are $\binom{N}{i}$ nodes of size $i$, making the total computation $\sum_{i=1}^{N}2^{i}\binom{N}{i} = 3^N-1$.

In Corollary \ref{coroll:improvement} we state that Algorithm \ref{alg:partition_function_trellis} is super-exponentially more efficient than brute force methods that consider every possible hierarchy. Their ratio is
{\small 
\begin{align}
{\it r}=\frac{(2N-3)!!}{3^N}=\frac{1}{2 \sqrt{\pi}} \bigg(\frac{2}{3}\bigg)^N \Gamma(N-1/2)
\end{align}
}%
with $\Gamma$ the gamma function. Thus, {\it r} presents a super-exponential growth in terms of $N$.
% \subsection{Additional Details on PAM50 Experiments}
% \label{app:genomics}
% The Pearson correlation coefficient was used for the clustering metric for the PAM50 data set experiments. 
% The correlation clustering input can be represented as a complete weighted graph, $G = (V,E)$, where each edge has weight $w_{uv} \in [-1,1], \forall (u,v) \in E$. 
% The goal is to construct a clustering of the nodes that maximizes the sum of positive within-cluster edge weights minus the sum of all negative across-cluster edge weights. 
% However, the correlations among subsampled \texttt{pam50} ($n=12$) data set are all positive. 
% To allow more balanced weights among vertices, we transform the weights using $w_{uv}' = \tan \left(w_{uv} \frac{\pi}{2} \right)$ and then subtracted the average of all $w_{uv}'$.

% The MAP tree shown in the main paper uses the correlation coefficient and the construction code had a small error.\sebastian{Is there a typo in the previous sentence?}
% With transformed weights, the MAP tree is shown Figure \ref{fig: map_tree}.
% \begin{figure}[h!]
%   \centering
%   \includegraphics[scale = 0.45]{figs/tangent_trellis_square.PNG}
%   \caption{MAP tree on the subsampled \texttt{pam50} ($n=12$) data set.}
%   \label{fig: map_tree}
% \end{figure}
% We note that the subtree rooted at $\{7,8,9,10,11,12\}$ is consistent.
% All of the correlation coefficients among this cluster are positive, so the optimal action is to split off the item with the smallest (positive) correlation coefficient.

\subsection{Jet Physics Background}
\label{jet-physics-background}
% A specific jet could result from several latent trees generated by the showering process. 
%Each latent path depends on the specific details of the $1 \rightarrow 2$ splittings. 
It is natural to represent a jet and the particular clustering history that gave rise to it as a binary tree, where the inner nodes represent each of the unstable particles and the leaves represent the jet constituents. This representation connects jets physics with natural language processing (NLP) and biology, which is exciting and was first suggested in \cite{Louppe:2017ipp}.  
%\subsubsection{Background}
%  \vspace{-0.4cm}
% \paragraph{Background}

Jets are among the most common objects produced at the Large Hadron Collider (LHC) at CERN, and a great amount of work has been done to develop techniques for a better treatment and understanding of them,  from both an experimental and theoretical point of view. In particular, determining the nature (type) of the initial unstable particle (the root of the binary tree), and its children and grandchildren that gave rise to a specific jet is essential in searches for new physics, as well as precision measurements of our current model of nature, i.e., the Standard Model of particle physics. 
In this context, it becomes relevant and interesting to study algorithms to cluster the jet constituents (leaves) into a binary tree and metrics to compare them. Being able to improve over the current techniques that attempt to invert the showering process to reconstruct the ground truth-level tree would assist in physics searches at the Large Hadron Collider. 

There are software tools called {\bf parton showers}, e.g., \href{http://home.thep.lu.se/Pythia/}{PYTHIA}, \href{https://herwig.hepforge.org/}{Herwig}, \href{https://sherpa.hepforge.org/trac/wiki}{Sherpa}, that encode a physics model for the simulation of jets that are produced at the LHC. 
Current algorithms used by the physics community to estimate the clustering history of a jet are domain-specific sequential recombination jet algorithms, called {\it generalized $k_t$ clustering algorithms} \cite{Cacciari:2008gp}, and they do not use these generative models. These algorithms sequentially cluster the jet constituents by locally choosing the pairing of nodes that minimizes a distance measure. Given a pair of nodes, this measure depends on the angular distance between their momentum vector and the value of this vector in the transverse direction with respect to the collision axis between the incoming beams of protons.

\begin{figure}[bhp]
\centering
\includegraphics[width=0.25\columnwidth]{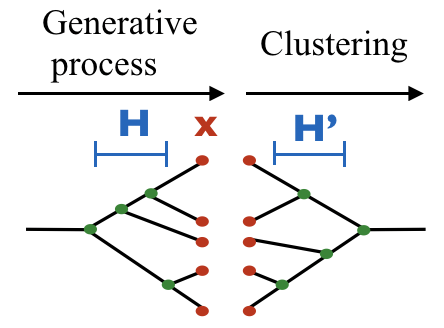}
\caption{\small{Schematic representation of the tree structure of a sample jet generated with Ginkgo and the clustered tree for some clustering algorithm. For a given algorithm, $z$ labels the different variables that determine the latent structure of the tree. The tree leaves $x$ are labeled in red and the inner nodes in green. 
}}
\label{fig:treestructure}
\end{figure}

Currently, generative models that implement the parton shower in full physics simulations are implicit models, i.e., they do not admit a tractable density.
%are a probabilistic model defined only in terms of the samples they generate. 
Extracting additional information that describes the features of the latent process is relevant to study problems where we aim to unify generation and inference, e.g inverting the generative model to estimate the clustering history of a jet. A schematic representation
of this approach is shown in Figure \ref{fig:treestructure}.

% \begin{SCfigure*}
% % \vspace{-0.2cm}
%     % \right
%     \caption{{\small Schematic representation of the tree structure of a sample jet generated with Ginkgo and the clustered tree for some clustering algorithm. For a given algorithm, $z$ labels the different variables that determine the latent structure of the tree. The tree leaves $x$ are labeled in red and the inner nodes in green. }}
%             \includegraphics[width=0.3\textwidth]{figs/JetsLatentStructurec.png}
% % \vspace{0.2cm}
%     \label{fig:treestructure}
%     % \vspace{-0.2cm}
% \end{SCfigure*}

% \vspace{-2cm}
% and approximate the latent structure of the generated jet based on the maximum likelihood estimate.
% \vspace{-0.15cm}
At present, it is very hard to access the joint likelihood in state-of-the-art parton shower generators in full physics simulations. Also, typical implementations of parton showers involve 
%kinematic reshuffling 
sampling procedures that destroy the analytic control of the joint likelihood.
Thus, to aid in machine learning (ML) research for jet physics, a python package for a toy generative model of a parton shower, called Ginkgo, was introduced in \cite{ToyJetsShowerPackage}.
Ginkgo has a tractable joint likelihood, and is as simple and easy to describe as possible but at the same time captures the essential ingredients of parton shower generators in full physics simulations. 
Within the analogy between jets and NLP, Ginkgo can be thought of as ground-truth parse trees with a known language model.
A python package with a pyro implementation of the model with few software dependencies is publicly available in \cite{ToyJetsShowerPackage}. 
 \vspace{-0.4cm}
 
 \subsection{Counting Trees}
 We count the total number of hierarchies\footnote{This gives a result matching exactly the formula $\exactNumClusterings$}. We implement a bottom-up approach and start by assigning a number of trees  $N = 1$ to to each cluster of one element. Then, given a parent cluster $\dataset_p$, we add the contribution $N^i_p$ ($N_p = \sum_i N^i_p$) of each possible pair $i$ of left and right children, $s_{\dataset_p} = \{\dataset_L, \dataset_R\}$, where $\dataset_L \cup \dataset_R = \dataset_p$ and $ \dataset_L\cap \dataset_R = \emptyset$. In particular, we obtain
\begin{align}\label{eq:treecounting}
N^i_p = N^i_{X_L} \cdot N^i_{X_R}
\end{align}
Thus $N_p$ is the number of possible trees of the sub-branch whose root node is $X_p$. We repeat the process until we reach the cluster of all elements $\dataset$.
 
 \subsection{Runtime Asymptotics Plots}
\begin{figure} 
    \centering
       \includegraphics[width=0.5\columnwidth]{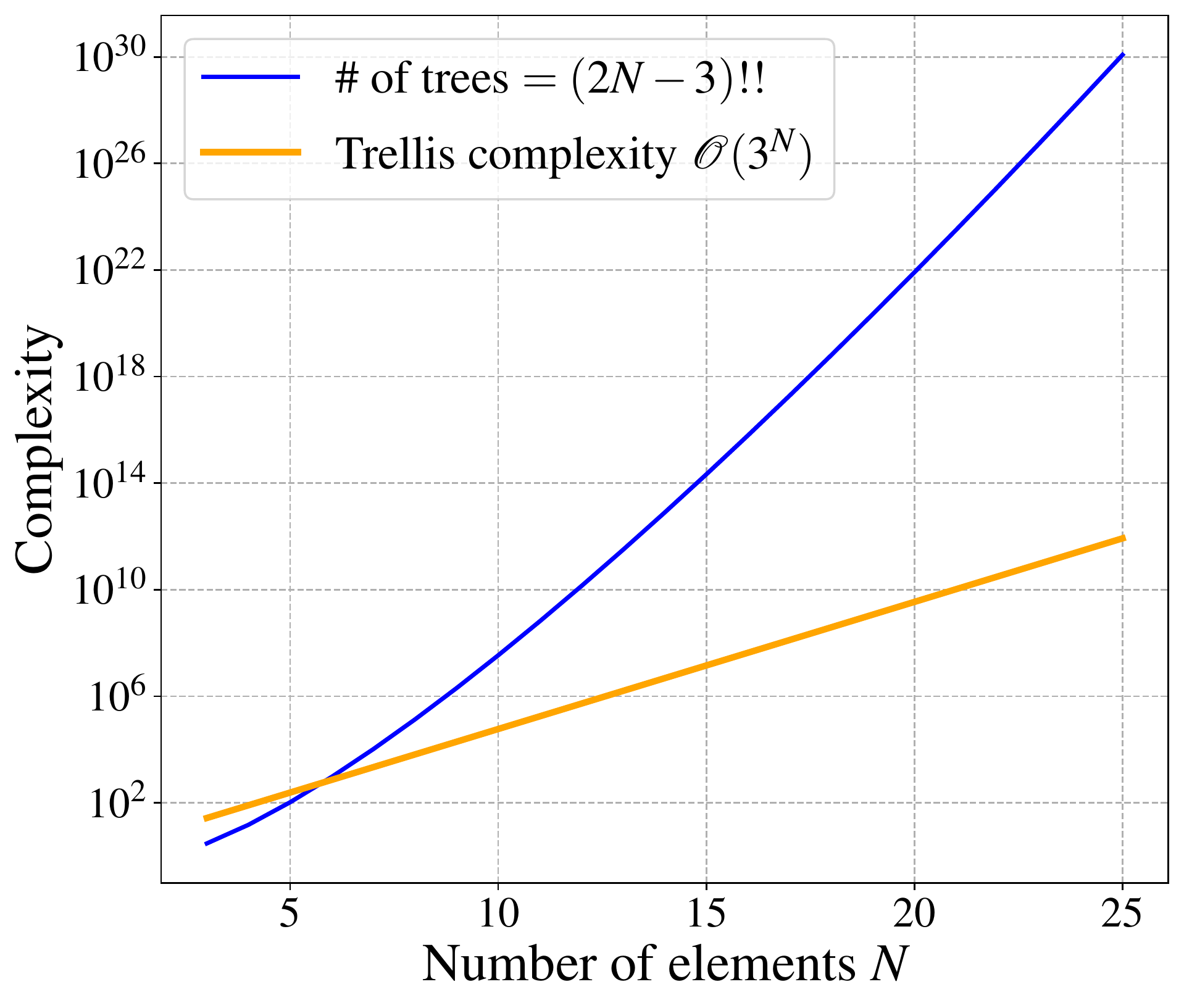}
        \caption{\small{Comparison of the complexity of the cluster trellis (orange) and the number of trees (blue) vs the number of elements of a dataset. We can see the super-exponential improvement (note the log scale on the y-axis) of Algorithm \ref{alg:partition_function_trellis} over brute force methods that consider every possible hierarchy, as introduced in Corollary \ref{coroll:improvement} and derived in Appendix $\S$ \ref{proof:Z-time-complexity}.}}
        \label{fig:time-asym}
 \end{figure}

 See Figure \ref{fig:time-asym} for a comparison of the number of trees vs the time complexity of the trellis algorithms for finding the partition function, MAP, and marginal values.
 
 \subsection{Description of Computer Architecture and Experimental Runtime}
%  \craig{I presume AIStats will be happy to see this?}
 When using a MacBook Pro with a 2.5 GHz Intel Core i5 processor with 8GB 1600 MHz DDR3 RAM to compute the MAP for the genetics experiments using the PAM dataset, it takes approximately 15 minutes to complete from start to finish (including data loading and result output).  When using this same machine to compute that MAP for Dasgupta cost on the given graph, it takes approximately 4 seconds to complete from start to finish (including data loading and result output).
 
When using a MacBook Pro with a 2.3 GHz Intel Core i9 processor with 16GB 2400 MHz DDR4 RAM to compute the MAP for the jet physics experiments it takes $5x10^{-2}$, 1.6 and 6.1 seconds to run the trellis on jets with 5, 9 and 10 leaves respectively.

\subsection{Sparse Trellis}\label{sparsetrellis}

\begin{figure}[ht]
    \centering
    \begin{minipage}{0.51\textwidth}
        \centering
        \includegraphics[width=\columnwidth]{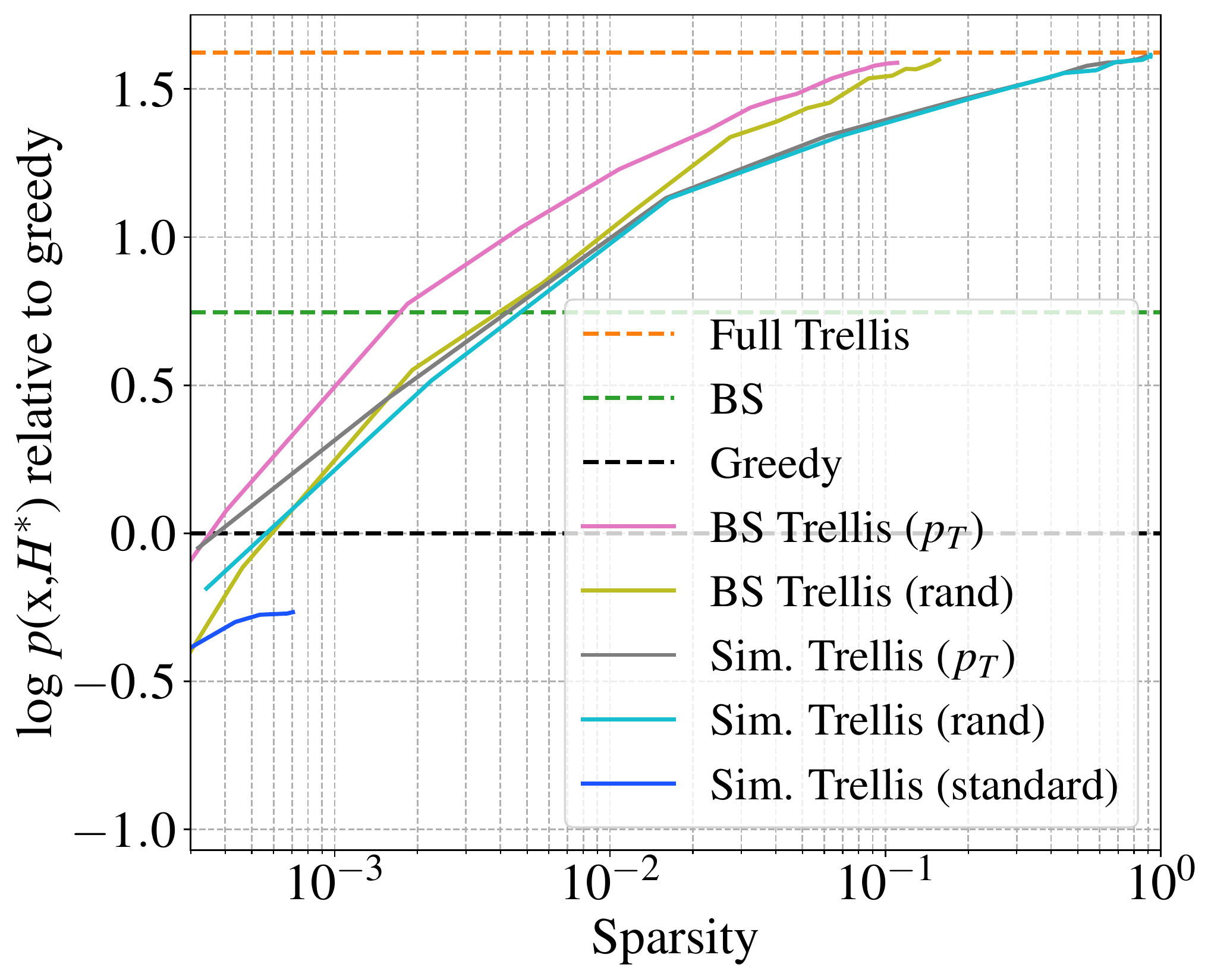}
        \caption{\small{Trellises MAP hierarchy log likelihood (values are relative to the greedy algorithm) vs their sparsity. Each value corresponds to the mean over 100 trees of a test dataset. We show the Simulator (Sim.) and the Beam Search (BS) trellises. 
% In both cases, we present the trellis obtained by ordering the leaves of the input trees in increasing norm of their momentum vector $\vec{p} \in \RR^3$ (see the probabilistic model description of section \ref{exp:jets} for more details). We compare sparse trellises with other orderings of the leaves in Appendix $\S$ \ref{sparsetrellis}.
% We add the values of the exact trellis, beam search and greedy algorithms. The BS trellis approaches the performance of the full one for a smaller sparsity index than the Sim. Trellis. 
% % Also, the results shown are obtained from running the sparse trellises on a different dataset than the one used to , after having being pre-built on a different one
% Also, the sparse trellises are pre-built and then run on new datasets (test), which is why BS performs better than BS trellis sometimes.}
We present the trellis obtained by ordering the leaves of the input trees in three different ways. First, in increasing norm of their momentum vector $\vec{p} \in \RR^3$ ($p_T$), see the probabilistic model description of section \ref{exp:jets} for more details. Second, leaves ordered randomly (rand). Third, leaves ordered by how they are accessed by traversing the trees (standard). Note that in this last case, we only show the Sim. trellis results as the BS trellis spans over sparsity indices values of $\mathcal{O}(10^{-5})$ and has a worse performance. 
We add the values of the full trellis, beam search and greedy algorithms. The BS trellis approaches the performance of the full one for a smaller sparsity index than the Sim. Trellis.
        }}
        \label{fig:MAPvsSparsity_full}
    \end{minipage}\hfill
    \begin{minipage}{0.45\textwidth}
        \centering
        \includegraphics[width=\columnwidth]{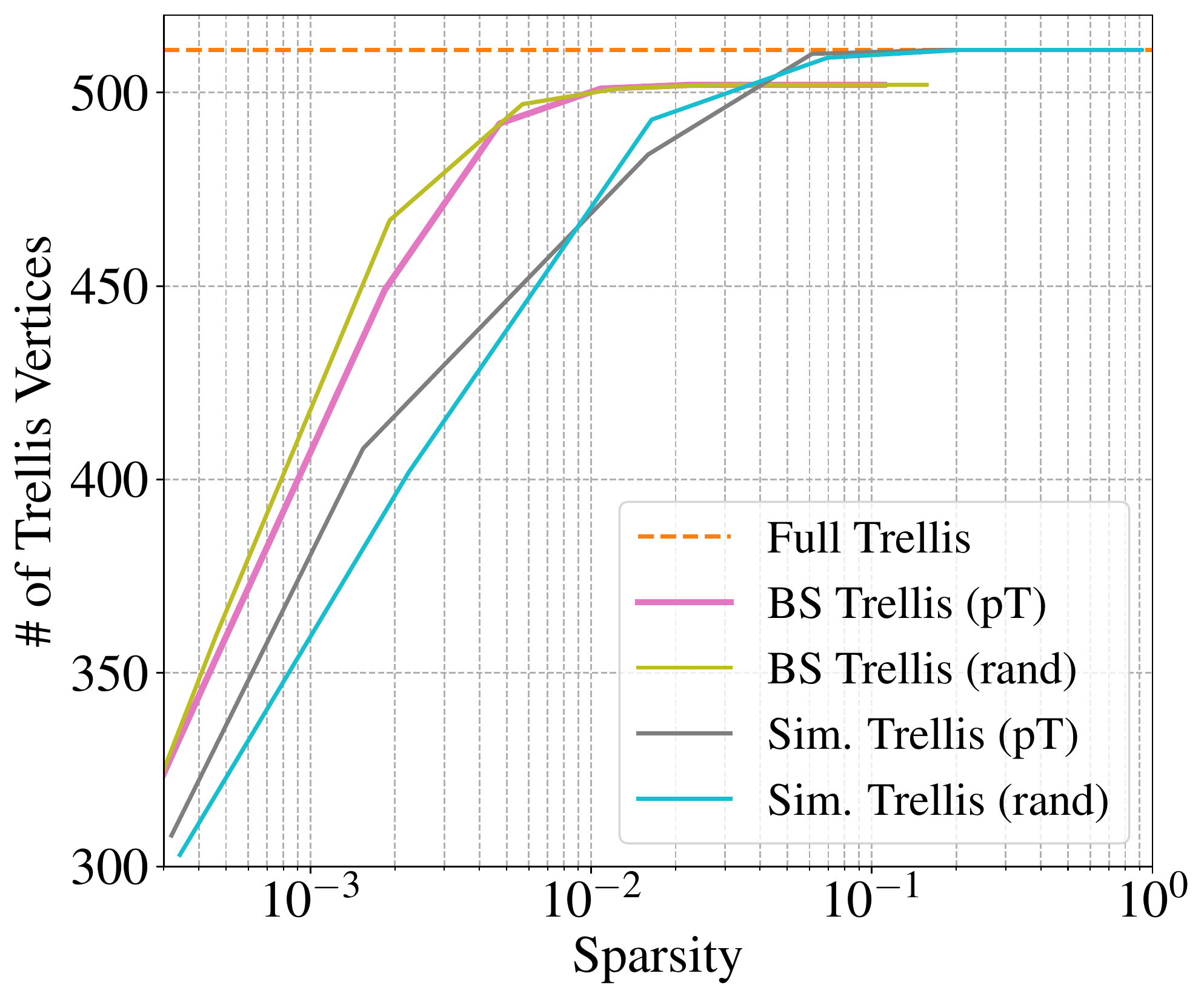}
        \caption{\small{Trellises number of vertices vs their sparsity. We show the Simulator (Sim.) and the Beam Search (BS) trellises. The leaves of the input trees are ordered in different ways, as explained in Figure \ref{fig:MAPvsSparsity_full}.
Sim. trellis saturates all the vertices below a sparsity of $\sim 0.1$ but the performance in Figure \ref{fig:MAPvsSparsity_full} keeps increasing. The reason is that we keep adding edges to existing vertices, thus realizing a greater number of trees. 
        }}
        \label{fig:NodesvsSparsity}
    \end{minipage}
\end{figure}

As mentioned in section \ref{sec:building}, there are different mappings for the ordering of the leaves of the input trees when building the sparse trellis, and the subset of hierarchies spanned by the trellis depends on this mapping. Specifically, two sub-hierarchies identical under some ordering of the leaves would contribute the same vertices and edges to the trellis. However, this could change by modifying the ordering, e.g. vertex $\{a,b\}$ could turn into vertices $\{a,b\}$ and $\{a,d\}$. Thus, the hierarchies over which the sparse trellis spans depend on the ordering of the leaves of the input trees that we use to build it.
We show in Figure \ref{fig:MAPvsSparsity_full} the performance of the sparse trellis to calculate the MAP values on a set of 100 Ginkgo jets with 9 leaves. Here we study the sparse trellises for more orderings of the leaves of the input trees than the ones shown in Figure \ref{fig:MAPvsSparsity}.
% We chose a dataset of 9 elements to be able to easily compare the performance of the sparse and full trellises. However, the sparse trellis can be applied to larger datasets. Even though beam search has a good performance for trees with a small number of leaves, we see that both sparse trellises quickly improve over beam search, with a sparsity index of only about 2\%. 
% % For these values of the sparsity index, the trellis is efficient and fast to run. 
% Both sparse trellises approach the performance of the exact one, but the BS trellis does it sooner.

Next, in Figure \ref{fig:NodesvsSparsity} we show the number of vertices added to the sparse trellis vs their sparsity (number of trees that they can realize over total possible number of trees). It is interesting to note that the sparsity depends not only on the number of vertices but also on their location in the trellis as well as the edges. Thus, we see that for the same number of vertices, there are different sparsity indices, depending on the building strategy.

\vspace{-.1cm}

\end{document}